\documentclass[prb,twocolumn,groupedaddress,shownopacs,amssymb, amsmath]{revtex4-2}
\usepackage{graphicx}

\usepackage[usenames, dvipsnames]{color}

\newcommand{\ii}{\mathrm{i}} % imaginary unit (roman), distinct from the Cartesian index i

\usepackage{hyperref}
\hypersetup{
    colorlinks=true,
    linkcolor=Blue,
    citecolor=Blue,
    filecolor=Blue,
    urlcolor=Blue,
    pdftitle={pair quantum geometry},
    }
\begin{document}

\title{Quantum geometry of collective pairing fluctuations \\
in the superfluid weight of multiband superconductors}
\author{R. N. Kalkan and M. Iskin}
\affiliation{
Department of Physics, Ko\c{c} University, Rumelifeneri Yolu,
34450 Sar\i yer, Istanbul, T\"urkiye
}

\date{\today}

\begin{abstract}

Beyond the frozen-pairing contribution, the superfluid weight of a
multiband superconductor contains a correction from the self-consistent
relaxation of the pairing amplitudes under a phase twist. We show that
this correction has a quantum-geometric origin, not in the space of Bloch
states but in the space of collective pairing fields. For an attractive
multiband Hubbard model at the mean-field level, the thermodynamic Hessian
governing the pairing response coincides with the static Gaussian
pair-fluctuation kernel at zero momentum at every temperature below
$T_c$. Since gauge invariance relates a phase twist to a finite
center-of-mass momentum of the pair field, the superfluid weight is set
by the small-momentum curvature of the eigenvalue branch that evolves from
the Goldstone mode, and the relaxation correction is a stiffness-weighted
quantum metric of the corresponding soft eigenvector on the manifold of
pairing configurations. The same construction yields the Ginzburg-Landau
gradient coefficient with the identical geometric correction, hence the
coherence-length tensor, and, with the low-frequency dynamics of the
kernel, the pair-mass tensor. Lieb-lattice calculations verify the
equivalence of the thermodynamic, fluctuation-kernel, and soft-mode
formulations and the independence of the superfluid weight from the
orbital embedding.

\end{abstract}
\maketitle

\section{Introduction}
\label{sec:intro}

The superfluid weight~\cite{fisher73, scalapino92} of a multiband superconductor
is determined not only by the dispersion of its energy bands but also by the
quantum geometry of their Bloch states~\cite{peotta15, liang17}. In particular,
interband processes can generate a geometric contribution that remains finite
even in flat-band systems, where the conventional contribution associated with
band dispersion is suppressed. This connection between quantum geometry and
superfluidity has motivated extensive interest in multiband and flat-band
superconductors~\cite{torma22, peotta23, yu25, liu25, gao25, jiang25, verma26,
kitamura26, stern26, tupitsyn26, berger26}.
Experiments on moir\'e graphene support this picture: the measured superfluid
stiffness of magic-angle twisted bilayer graphene exceeds the conventional
band-dispersion estimate by an order of magnitude, consistent with a dominant
quantum-geometric contribution~\cite{Tian23,Tanaka25}. Its temperature and
current dependence point to a nodal gap~\cite{Tanaka25,Banerjee25}, while short
coherence lengths comparable to the interparticle distance and gap-to-$T_c$
ratios far above the BCS value place these systems in the strong-coupling
regime~\cite{Cao18,Park21,Oh21,Kim22}.

The geometry of the underlying Bloch states, however, is not the only source of
corrections to the superfluid weight. In a multiorbital superconductor the
pairing amplitudes on different orbitals are independent degrees of freedom,
so a phase twist can change their magnitudes and relative phases even when
the equilibrium pairing is spatially uniform; a superfluid weight computed
with the amplitudes held fixed misses this
response~\cite{chan22, huhtinen22, hu24, buthenhoff25, prijon26}. Huhtinen
\textit{et al.}~\cite{huhtinen22} identified the resulting relaxation
correction and expressed it through the order-parameter response and the
Hessian of the mean-field thermodynamic potential, which provides a
general formulation of the self-consistent superfluid weight. Less
explicit in that formulation is its connection to collective pairing
fluctuations and to the momentum dependence of their
eigenmodes~\cite{keskiner26, sun26, simon26}, even though a phase twist
couples directly to the finite-momentum pair field while the
thermodynamic Hessian describes only its static response at zero
momentum.

In this work, we connect these two descriptions directly. For an
attractive multiband Hubbard model at the mean-field saddle-point level,
we show that the Hessian of the thermodynamic potential with respect to
the complex pairing fields coincides with the Gaussian pair-fluctuation
kernel at zero momentum and frequency, at every temperature below $T_c$.
Since gauge invariance maps a uniform phase twist onto a finite
center-of-mass momentum $\mathbf Q$ of the pair field, the
self-consistent superfluid weight follows from the small-momentum
curvature of the eigenvalue branch of this kernel that evolves from the
Goldstone mode. The relaxation correction arises because at finite
$\mathbf Q$ the Goldstone mode couples to the massive amplitude and
relative-phase channels; integrating them out yields a Schur complement
whose subtracted term is a stiffness-weighted quantum metric of the soft
eigenvector on the manifold of pairing configurations. This collective
metric is distinct from the Bloch quantum metric, and it vanishes when
the Goldstone mode decouples to linear order in $\mathbf Q$. At $T_c$
the same projection onto the critical pairing channel gives the
Ginzburg-Landau (GL) gradient coefficient with the same geometric
correction, and hence the coherence-length tensor~\cite{keskiner26},
whereas the pair mass additionally requires the low-frequency dynamics
of the kernel. For the Lieb lattice we verify numerically that the
thermodynamic, fluctuation-kernel, and soft-mode formulations agree,
that the frozen-pairing and relaxation contributions are separately
embedding dependent while their sum is not, and that in the disconnected
limit the relaxation correction cancels the spurious frozen-pairing
stiffness~\cite{huhtinen22}.

The paper is organized as follows. In Sec.~\ref{sec:thermo}, we develop the
thermodynamic formulation of the superfluid weight and isolate the
pairing-relaxation correction. In Sec.~\ref{sec:effective_action}, we
connect it to Gaussian pair fluctuations and develop the soft-mode and
quantum-geometric description. In Sec.~\ref{sec:supfwe}, we assemble the
resulting superfluid weight and discuss the conditions under which the
relaxation correction vanishes. In Sec.~\ref{sec:Tc}, we extend the
construction to the critical pairing channel and derive the
coherence-length and pair-mass tensors. In Sec.~\ref{sec:validity}, we
summarize the assumptions and scope of the approach, and in
Sec.~\ref{sec:numerics}, we present the Lieb-lattice benchmark. In
Sec.~\ref{sec:conclusion}, we conclude the paper. The appendices derive
the kernel-Hessian relation in an alternative way, examine the kernel 
under a change of orbital embedding, and establish the gauge relations 
between phase-twist and pair-momentum derivatives.

\section{Thermodynamic formulation}
\label{sec:thermo}

We consider an attractive Hubbard model on a translationally invariant lattice
with $N_S$ sublattice sites (orbitals) per unit cell,
\begin{align}
    H=&\sum_{ll'}\sum_{SS'\sigma}t_{SS'}^\sigma(\mathbf R_l-\mathbf R_{l'})
       c_{lS\sigma}^\dagger c_{l'S'\sigma}-\mu\sum_{lS\sigma}n_{lS\sigma}
    \nonumber \\
    &-U\sum_{lS}n_{lS\uparrow}n_{lS\downarrow},\qquad U>0,
\label{eq:H_real}
\end{align}
where $c_{lS\sigma}^\dagger$ and $c_{lS\sigma}$ create and annihilate a fermion
of spin $\sigma$ on orbital $S$ of unit cell $l$, and
$
n_{lS\sigma}=c_{lS\sigma}^\dagger c_{lS\sigma}.
$
The index $l = 1, \ldots, N_c$ labels the unit cells, located at $\mathbf R_l$,
and $S = 1, \ldots, N_S$ labels the orbitals within a unit cell; the indices
$i$ and $j$ are reserved throughout for Cartesian directions, and the
imaginary unit is written in roman type as $\ii$. We denote the
position of orbital $S$ relative to the unit-cell origin by $\mathbf x_S$, so
that its real-space position is $\mathbf r_{lS}=\mathbf R_l+\mathbf x_S$.
Translational invariance implies that the hopping amplitudes depend only on
the unit-cell displacement
$
\mathbf R = \mathbf R_l-\mathbf R_{l'},
$
while Hermiticity requires
$
t_{SS'}^\sigma(\mathbf R)=[t_{S'S}^\sigma(-\mathbf R)]^*.
$
The last term in Eq.~\eqref{eq:H_real} describes an on-site attraction of strength $U$ between
opposite-spin fermions, taken for simplicity to be identical on every orbital.
Throughout, we set $\hbar=k_B=1$ and work in the grand-canonical ensemble at
fixed chemical potential $\mu$.

We use the Fourier transformation
$
c_{lS\sigma}=\frac{1}{\sqrt{N_c}}\sum_{\mathbf k}
e^{\ii\mathbf k\cdot\mathbf r_{lS}}c_{\mathbf kS\sigma},
$
where the orbital positions appear explicitly, and $\mathbf k$ runs over the
first Brillouin zone (BZ) where $\sum_{\mathbf k\in\mathrm{BZ}}1=N_c$.
The single-particle part of Eq.~\eqref{eq:H_real} is then described by the
$N_S\times N_S$ Bloch matrix
$
[h_{\mathbf k\sigma}]_{SS'}=\sum_{\mathbf R}
t_{SS'}^\sigma(\mathbf R)e^{-\ii\mathbf k\cdot(\mathbf R+\mathbf x_S-\mathbf x_{S'})}.
$
The Hermiticity condition on the real-space hopping amplitudes guarantees
$
\mathbf h_{\mathbf k\sigma}=\mathbf h_{\mathbf k\sigma}^\dagger
$
for every $\mathbf k$. The momentum-space Hamiltonian then takes the form
\begin{equation}
    H=\sum_{\mathbf k\sigma}\mathbf c_{\mathbf k\sigma}^\dagger
    \left(\mathbf h_{\mathbf k\sigma}-\mu \mathbf{1} \right)\mathbf c_{\mathbf k\sigma}
    -\frac{U}{N_c}\sum_{S \mathbf Q}b_{S\mathbf Q}^\dagger b_{S\mathbf Q},
    \label{eq:H_momentum}
\end{equation}
where
$
\mathbf c_{\mathbf k\sigma}=(c_{\mathbf k1\sigma},\ldots,
c_{\mathbf kN_S\sigma})^{\mathrm T},
$
with $\mathrm{T}$ denoting transposition, $\mathbf 1$ is the identity matrix
of the appropriate dimension (here $N_S\times N_S$), and
$
b_{S\mathbf Q}^\dagger=\sum_{\mathbf k}
c_{\mathbf k+\mathbf Q,S\uparrow}^\dagger c_{-\mathbf k,S\downarrow}^\dagger
$
creates an on-site singlet pair on orbital $S$ with center-of-mass momentum
$\mathbf Q$. Note that the orbital phases $e^{\pm \ii\mathbf k\cdot\mathbf x_S}$
carried by the Fourier convention cancel in the interaction by momentum
conservation, so that the embedding $\{\mathbf x_S\}$ enters only through
$\mathbf h_{\mathbf k\sigma}$; Appendix~\ref{app:embedding} discusses how
the momentum-space representation transforms under a change of embedding
representing the same real-space Hamiltonian. Since the interaction is
diagonal in $S$, the pairing channels are labeled  by the orbital index
and the pairing field has $N_S$ components. At the mean-field level,
we decouple the interaction in the pairing channel by introducing
the pairing amplitudes $\Delta_S$, which at self-consistency are given by
\begin{equation}
    \Delta_S=-U\langle c_{lS\downarrow}c_{lS\uparrow}\rangle,
    \label{eq:Delta_def}
\end{equation}
independent of the unit-cell index $l$ by translational invariance.
We collect them into the pairing vector
$
\boldsymbol{\Delta}=(\Delta_1,\ldots,\Delta_{N_S})^{\mathrm T},
$
with
$
|\boldsymbol{\Delta}|^2=\boldsymbol{\Delta}^\dagger\boldsymbol{\Delta}
=\sum_S|\Delta_S|^2,
$
and treat $\boldsymbol{\Delta}$ as a variational parameter until it is fixed
by the saddle-point condition below; its self-consistent value is denoted
$\boldsymbol{\Delta}_0$, with components $\Delta_{0S}$.
Dropping the term quadratic in the fluctuations about these averages replaces
the interaction in Eq.~\eqref{eq:H_momentum} by
\begin{equation}
    \sum_{\mathbf kS}\big(
    \Delta_Sc_{\mathbf kS\uparrow}^\dagger c_{-\mathbf kS\downarrow}^\dagger
    +\Delta_S^*c_{-\mathbf kS\downarrow}c_{\mathbf kS\uparrow}
    \big)
    +N_c\frac{|\boldsymbol{\Delta}|^2}{U},
    \label{eq:pairing_term}
\end{equation}
where the last term is the usual counterterm.

To determine the superfluid weight, we impose a uniform phase twist $\mathbf q$
on the pairing field,
$
\Delta_S\rightarrow\Delta_S e^{2\ii\mathbf q\cdot\mathbf r_{lS}},
$
where the factor of $2$ reflects the charge-two nature of the pair field. A gauge
transformation removes this spatial phase from the pairing field and transfers
the twist to the fermionic kinetic term, so that a pair is formed by
$(\mathbf k+\mathbf q,\uparrow)$ and $(\mathbf q-\mathbf k,\downarrow)$ and the
problem is again uniform~\cite{fisher73, taylor06}.
In this gauge, we introduce the Nambu spinor
$
\mathbf C(\mathbf k,\mathbf q)=(\mathbf c_{\mathbf k+\mathbf q,\uparrow},
\mathbf c^{\dagger}_{\mathbf q-\mathbf k,\downarrow})^{\mathrm{T}}.
$
The mean-field Hamiltonian and the corresponding Bogoliubov-de Gennes (BdG)
Hamiltonian then take the form
\begin{align}
    H_{0} (\mathbf q)
    &=
    \sum_{\mathbf k}
    \mathbf C^{\dagger}(\mathbf k, \mathbf q)
    \mathbf h_{\mathrm{BdG}} (\mathbf k,\mathbf q)
    \mathbf C(\mathbf k, \mathbf q)
    \nonumber \\
    &\quad
    +\sum_{\mathbf k}\operatorname{tr}
    \left(
        \mathbf h_{\mathbf q-\mathbf k,\downarrow}-\mu \mathbf{1}
    \right)
    +N_c\frac{|\boldsymbol{\Delta}|^2}{U},
    \nonumber \\
    \mathbf h_{\mathrm{BdG}} (\mathbf k,\mathbf q)
    &=
    \begin{bmatrix}
        \mathbf h_{\mathbf k+\mathbf q,\uparrow}-\mu \mathbf{1}
        &
        \operatorname{diag}(\boldsymbol{\Delta})
        \\
        \operatorname{diag}(\boldsymbol{\Delta})^\dagger
        &
        -\left(
        \mathbf h_{\mathbf q-\mathbf k,\downarrow}-\mu \mathbf{1}
        \right)^{\mathrm T}
    \end{bmatrix},
    \label{eq:HMF_HBdG}
\end{align}
where, here and throughout, $\operatorname{tr}$ denotes the matrix trace over
the orbital or Nambu-orbital space on which its argument acts (here the
$N_S$-dimensional orbital space).
The mean-field partition function is defined as
$
\mathcal Z_{0}(\mathbf q)
=\operatorname{Tr}_{\mathrm{Fock}}e^{-\beta H_{0}(\mathbf q)},
$
with $\beta=1/T$, where $\operatorname{Tr}_{\mathrm{Fock}}$ runs over the
full Fock space, and the thermodynamic potential per unit cell as
$
\Omega_0(\mathbf q,\boldsymbol{\Delta},\boldsymbol{\Delta}^*)
=-\frac{T}{N_c}\ln\mathcal Z_{0}(\mathbf q).
$
Diagonalizing $\mathbf h_{\mathrm{BdG}}(\mathbf k,\mathbf q)$,
with eigenvalues $E_p(\mathbf k,\mathbf q)$, the thermodynamic potential becomes
\begin{align}
    \Omega_0
    (\mathbf q,\boldsymbol{\Delta},\boldsymbol{\Delta}^*)
    =&
    \frac{|\boldsymbol{\Delta}|^2}{U}
    +\frac{1}{N_c}\sum_{\mathbf k}
    \Big\{
    \operatorname{tr}
    \left(
        \mathbf h_{\mathbf q-\mathbf k,\downarrow}
        -\mu \mathbf{1}
    \right)
    \nonumber \\
    &
    -T\sum_{p=1}^{2N_S}\ln \left[
        1+e^{-E_p(\mathbf k,\mathbf q)/T}
    \right]
    \Big\}.
\label{eq:Omega_0}
\end{align}
The constant trace term loses its apparent $\mathbf q$ dependence once $\mathbf k$
runs over the BZ, so that the twist enters Eq.~\eqref{eq:Omega_0}
only through the quasiparticle spectrum.

At fixed $\mathbf q$, the self-consistent pairing vector is determined by the
saddle-point equation
\begin{equation}
    \frac{\partial\Omega_0(\mathbf q,\boldsymbol{\Delta},\boldsymbol{\Delta}^*)}
    {\partial\boldsymbol{\Delta}^*}=\mathbf 0 ,
    \label{eq:gap}
\end{equation}
i.e.\ $\partial\Omega_0/\partial\Delta_S^*=0$ for every $S$, where
$\boldsymbol{\Delta}$ and $\boldsymbol{\Delta}^*$ are treated as independent
variables. By the Hellmann-Feynman theorem applied to Eq.~\eqref{eq:pairing_term},
$
\partial\Omega_0/\partial\Delta_S^*
=\Delta_S/U+\langle c_{lS\downarrow}c_{lS\uparrow}\rangle,
$
so that Eq.~\eqref{eq:gap} is precisely the self-consistency
condition~\eqref{eq:Delta_def}; its complex conjugate,
$
\partial\Omega_0/\partial\boldsymbol{\Delta}=\mathbf 0,
$
carries no further information.
The imposed twist fixes only the winding $e^{2\ii\mathbf q\cdot\mathbf r_{lS}}$ of
the pairing field; the underlying amplitudes are not held fixed, and
Eq.~\eqref{eq:gap} has to be solved at each $\mathbf q$. Its solution
$\boldsymbol{\Delta}_0(\mathbf q)$ generally differs from
$\boldsymbol{\Delta}_0(\mathbf 0)$ not merely by an overall phase but in the
ratios and the relative phases of the $N_S$ components~\cite{chan22}.
Substituting this solution back into the thermodynamic potential defines the
potential along the self-consistent manifold,
$
\Omega(\mathbf q)=\Omega_0\left(\mathbf q,\boldsymbol{\Delta}_0(\mathbf q),
\boldsymbol{\Delta}_0^*(\mathbf q)\right).
$

The superfluid-weight tensor per unit volume is obtained from the curvature of
the thermodynamic potential along the self-consistent solution,
\begin{equation}
    V[D_s]_{ij}
    =
    N_c
    \frac{\mathrm{d}^2\Omega(\mathbf q)}{\mathrm{d}q_i\,\mathrm{d}q_j}
    \bigg|_{\mathbf q=\mathbf 0},
    \label{eq:D_s}
\end{equation}
where $V$ is the system volume. Because $\boldsymbol{\Delta}_0(\mathbf q)$
depends on the twist, evaluating the total derivative in Eq.~\eqref{eq:D_s}
requires the expansion of $\Omega_0$ to second order in the pairing field.
Because the superconductor breaks the global
$U(1)$ symmetry, a variation $\delta\boldsymbol{\Delta}$ couples to
$\delta\boldsymbol{\Delta}$ as well as to $\delta\boldsymbol{\Delta}^*$, and it
is convenient to work in doubled complex coordinates. For a general pairing
vector we write
$
\boldsymbol{\Psi}=(\boldsymbol{\Delta},\boldsymbol{\Delta}^*)^\mathrm{T}
$
and
$
\Omega_0(\mathbf q,\boldsymbol{\Psi})\equiv
\Omega_0(\mathbf q,\boldsymbol{\Delta},\boldsymbol{\Delta}^*);
$
the self-consistent solution is
$
\boldsymbol{\Psi}_0(\mathbf q) = (\boldsymbol{\Delta}_0(\mathbf q),
\boldsymbol{\Delta}_0^*(\mathbf q))^\mathrm{T},
$
a variation about it is
$
\delta\boldsymbol{\Psi} = (\delta\boldsymbol{\Delta},
\delta\boldsymbol{\Delta}^*)^\mathrm{T},
$
and throughout we abbreviate the untwisted saddle as
$
\boldsymbol{\Delta}_0\equiv\boldsymbol{\Delta}_0(\mathbf 0)
$
and
$
\boldsymbol{\Psi}_0\equiv\boldsymbol{\Psi}_0(\mathbf 0).
$
Expanding at fixed twist about the stationary point gives
\begin{equation}
    \Omega_0 \left(
    \mathbf q,\boldsymbol{\Psi}_0(\mathbf q)+\delta\boldsymbol{\Psi}
    \right)
    =
    \Omega(\mathbf q)
    +\mathbf g^\dagger(\mathbf q)\delta\boldsymbol{\Psi}
    +\frac{1}{2}\delta\boldsymbol{\Psi}^\dagger
    \boldsymbol{\mathcal H}(\mathbf q)\delta\boldsymbol{\Psi},
    \label{eq:expansion}
\end{equation}
up to second order in $\delta \boldsymbol{\Psi}$, where
$
\mathbf g(\mathbf q,\boldsymbol{\Psi})=(\partial\Omega_0/\partial\boldsymbol{\Delta}^*,
\partial\Omega_0/\partial\boldsymbol{\Delta})^{\mathrm T}
$
is the gradient of $\Omega_0$ at an arbitrary pairing vector $\boldsymbol{\Psi}$ and
$
\mathbf g(\mathbf q)\equiv\mathbf g(\mathbf q,\boldsymbol{\Psi}_0(\mathbf q))
$
is its on-shell value, which collects Eq.~\eqref{eq:gap} and its conjugate and
therefore vanishes identically in $\mathbf q$,
and the Hessian is the $2N_S \times 2N_S$ matrix of second derivatives
with the same block ordering as $\boldsymbol{\Psi}$,
\begin{equation}
    \boldsymbol{\mathcal H}(\mathbf q)
    =
    \begin{pmatrix}
        \dfrac{\partial^2\Omega_0}
        {\partial\boldsymbol{\Delta}^*\partial\boldsymbol{\Delta}^{\mathrm T}}
        &
        \dfrac{\partial^2\Omega_0}
        {\partial\boldsymbol{\Delta}^*\partial\boldsymbol{\Delta}^{\dagger}}
        \\[4mm]
        \dfrac{\partial^2\Omega_0}
        {\partial\boldsymbol{\Delta}\partial\boldsymbol{\Delta}^{\mathrm T}}
        &
        \dfrac{\partial^2\Omega_0}
        {\partial\boldsymbol{\Delta}\partial\boldsymbol{\Delta}^{\dagger}}
    \end{pmatrix}
    \equiv
    \begin{pmatrix}
        \boldsymbol{\mathcal A} & \boldsymbol{\mathcal C}\\
        \boldsymbol{\mathcal C}^* & \boldsymbol{\mathcal A}^*
    \end{pmatrix},
\label{eq:H_blocks}
\end{equation}
all evaluated at $\boldsymbol{\Psi}_0(\mathbf q)$. The Hessian depends on
$\mathbf q$ both through the twisted Hamiltonian and through
$\boldsymbol{\Delta}_0(\mathbf q)$.
Because $\Omega_0$ is real, its four blocks are not independent:
$
\boldsymbol{\mathcal{A}}=\boldsymbol{\mathcal{A}}^{\dagger}
$
and
$
\boldsymbol{\mathcal{C}}=\boldsymbol{\mathcal{C}}^{\mathrm{T}},
$
so that $\boldsymbol{\mathcal{H}}(\mathbf{q})$ is Hermitian and its
quadratic form on variations of the form
$
(\delta\boldsymbol{\Delta},\delta\boldsymbol{\Delta}^*)^{\mathrm T}
$
is the real second variation of $\Omega_0$. At a stable saddle it is
positive semidefinite, $\boldsymbol{\mathcal H}(\mathbf q)\succeq0$,
and its spectral decomposition may be used freely.

Differentiating the gap equation with respect to the phase twist relates the
response of the pairing vector to this Hessian. Denoting the response by
$
\mathbf F_i=\partial_{q_i}\boldsymbol{\Psi}_0(\mathbf q)|_{\mathbf q=\mathbf 0}
=\left(\partial_{q_i}\boldsymbol{\Delta}_0(\mathbf q),
\partial_{q_i}\boldsymbol{\Delta}_0^*(\mathbf q)\right)^{\mathrm T}
|_{\mathbf q=\mathbf 0}
$
and the mixed derivative by
$
\mathbf B_i=\partial_{q_i}\mathbf g(\mathbf q,\boldsymbol{\Psi})
|_{\mathbf q=\mathbf 0,\boldsymbol{\Psi}=\boldsymbol{\Psi}_0},
$
i.e.\ the explicit $\mathbf q$ derivative of the gradient at fixed
pairing vector, which is not the derivative of $\mathbf g(\mathbf q)$
since the latter vanishes identically, the total derivative of
$
\mathbf g(\mathbf q,\boldsymbol{\Psi}_0(\mathbf q))=\mathbf 0
$
gives
\begin{equation}
    \boldsymbol{\mathcal{H}}(\mathbf{0})\mathbf F_i=-\mathbf B_i.
    \label{eq:hfc}
\end{equation}
Global $U(1)$ invariance,
$
\Omega_0(\mathbf q,e^{\ii\phi}\boldsymbol{\Delta},
e^{-\ii\phi}\boldsymbol{\Delta}^*)
=\Omega_0(\mathbf q,\boldsymbol{\Delta},\boldsymbol{\Delta}^*),
$
fixes the pairing vector only up to a common phase. For an infinitesimal global
phase rotation of the untwisted saddle, the corresponding variation is
$
\delta\boldsymbol{\Psi}=\phi\mathbf u_0,
$
where
$
\mathbf u_0=(\ii\boldsymbol{\Delta}_0,-\ii\boldsymbol{\Delta}_0^*)^{\mathrm T}.
$
Inserting $\delta\boldsymbol{\Psi}$ into Eq.~\eqref{eq:expansion} and
using $\mathbf g=\mathbf 0$ gives
\begin{equation}
    \boldsymbol{\mathcal H}(\mathbf 0)\mathbf u_0=\mathbf 0,
    \quad
    \langle\mathbf u_0|\mathbf u_0\rangle=2|\boldsymbol{\Delta}_0|^2 ,
    \label{eq:u0}
\end{equation}
i.e.\ the global phase rotation is a zero mode of the Hessian~\cite{huhtinen22}.
This is the thermodynamic form of the Goldstone theorem, and $\mathbf u_0$ reappears
in Sec.~\ref{sec:below_Tc} as the Goldstone direction of the pair-fluctuation
kernel. The component of $\mathbf F_i$ along $\mathbf u_0$ is left undetermined
by Eq.~\eqref{eq:hfc}, reflecting the freedom to choose the global phase, but it
drops out of every physical quantity below.

The first total derivative of $\Omega(\mathbf q)$ equals the partial derivative
at frozen pairing field, because the implicit contributions vanish by the gap
equation. This is not the case for the second derivative, since one then
differentiates the gradient of $\Omega_0$ with respect to $\mathbf q$, which the
gap equation does not eliminate. Using the chain rule together with
Eq.~\eqref{eq:hfc}, we obtain
\begin{equation}
    V[D_s]_{ij}
    =
    N_c
    \frac{\partial^2\Omega_0}
    {\partial q_i\partial q_j}
    \bigg|_{\mathbf q=\mathbf 0,
    \boldsymbol{\Psi}=\boldsymbol{\Psi}_0}
    -
    N_c \mathbf F_i^\dagger
    \boldsymbol{\mathcal H}(\mathbf 0)\mathbf F_j.
    \label{eq:Ds_thermo}
\end{equation}
The first term is the curvature evaluated at frozen pairing amplitudes,
$
V[D_{\mathrm{frozen}}]_{ij}
    \equiv
    N_c
    \frac{\partial^2\Omega_0}
    {\partial q_i\partial q_j}
    \big|_{\mathbf q=\mathbf 0,
    \boldsymbol{\Psi}=\boldsymbol{\Psi}_0} ,
$
which contains the usual band-dispersion and Bloch-geometric contributions. The
second term accounts for the self-consistent relaxation of the pairing field
under the phase twist: as the twist is switched on, the pairing vector
rearranges as
$
\boldsymbol{\Psi}_0(\mathbf q)
=\boldsymbol{\Psi}_0+\sum_iq_i\mathbf F_i+O(\mathbf{q}^2),
$
and the system lowers its energy by doing so. Since
$\boldsymbol{\mathcal H}(\mathbf 0)\succeq0$ at a
stable saddle, the quadratic form
$
\mathbf F_i^\dagger\boldsymbol{\mathcal H}(\mathbf 0)\mathbf F_j
$
is positive semidefinite, and it enters Eq.~\eqref{eq:Ds_thermo} with
a negative sign. The relaxation therefore lowers the superfluid weight,
$\mathbf D_s\preceq\mathbf D_{\mathrm{frozen}}$~\cite{huhtinen22}.
Equation~\eqref{eq:Ds_thermo} also has a variational reading: applying the
same chain rule along an arbitrary smooth path
$\tilde{\boldsymbol{\Psi}}(\mathbf q)$ through the saddle, with tangent
$\tilde{\mathbf F}_i$, gives
$
N_c\,\mathrm{d}^2\Omega_0(\mathbf q,\tilde{\boldsymbol{\Psi}}(\mathbf q))
/\mathrm{d}q_i\,\mathrm{d}q_j |_{\mathbf 0}
=V[D_s]_{ij}
+N_c(\tilde{\mathbf F}_i-\mathbf F_i)^\dagger
\boldsymbol{\mathcal H}(\mathbf 0)(\tilde{\mathbf F}_j-\mathbf F_j),
$
so that every trial path, the frozen one included, overestimates the
superfluid weight.

A corresponding relaxation correction was obtained in Ref.~\cite{huhtinen22} at
fixed particle number using a gauge-fixed parametrization of the order
parameter. The two treatments differ in the ensemble and in the gauge, and
neither difference affects the content of Eq.~\eqref{eq:Ds_thermo}.
At fixed particle number one differentiates the free energy
$F=\Omega+\mu N$, so
that $\mu=\mu(\mathbf q)$ becomes an additional self-consistent variable
determined by $N=-\partial\Omega/\partial\mu$~\cite{taylor06}.
Treating it in the same way as
the pairing amplitudes above, with $\Omega(\mathbf q,\mu)$ the gap-relaxed
potential of Eq.~\eqref{eq:Ds_thermo} at fixed $\mu$, gives
$
\frac{\mathrm{d}^2F}{\mathrm{d}q_i\,\mathrm{d}q_j}\big|_{N} 
= \frac{\mathrm{d}^2\Omega}{\mathrm{d}q_i\,\mathrm{d}q_j}\big|_{\mu}
-\frac{(\partial_{q_i}\partial_\mu\Omega)(\partial_{q_j}\partial_\mu\Omega)}
 {\partial_\mu^2\Omega}.
$
The second term is the relaxation of the chemical potential. It carries the
opposite sign to the gap relaxation, because
$\partial_\mu^2\Omega=-\partial N/\partial\mu<0$, so that the fixed-$N$
stiffness is never smaller than the fixed-$\mu$ one. The mixed derivative
$\partial_{q_i}\partial_\mu\Omega$ vanishes at $\mathbf q=\mathbf 0$ whenever
$\Omega(\mathbf q)=\Omega(-\mathbf q)$, for instance in the presence of
time-reversal or inversion symmetry, in which case the two ensembles give the
same superfluid weight~\cite{huhtinen22}. As for the gauge,
Ref.~\cite{huhtinen22} fixes the global phase by choosing one
pairing amplitude to be real and positive, which removes the associated phase
degree of freedom and renders the reduced Hessian invertible. Here we retain the
full doubled complex pairing space, including both amplitude and phase
variations. Consequently $\boldsymbol{\mathcal H}(\mathbf 0)$ contains the zero
mode $\mathbf u_0$ of Eq.~\eqref{eq:u0} and is not invertible in the full
pairing space, which Eq.~\eqref{eq:Ds_thermo} never requires.

\section{Effective-action formulation}
\label{sec:effective_action}

We now turn to the effective-action description of pairing fluctuations.
Here no twist is imposed: spatial variations enter through the momentum
dependence of the fluctuations about the untwisted saddle, and a fermionic
twist $\mathbf q$ corresponds, by a gauge transformation, to a pairing field
of center-of-mass momentum $\mathbf Q=2\mathbf q$ (Appendix~\ref{app:Bi}).
This is what allows both terms of Eq.~\eqref{eq:Ds_thermo} to be read off
from the momentum dependence of the Gaussian fluctuation kernel.

\subsection{Effective action for the pairing field}
\label{sec:effaction}

We start from the coherent-state path-integral representation of the partition
function,
$
\mathcal Z=\int\mathcal D[\bar{\boldsymbol{\psi}},\boldsymbol{\psi}]
e^{-\mathcal S[\bar{\boldsymbol{\psi}},\boldsymbol{\psi}]},
$
where
$
\boldsymbol{\psi}_{k\sigma}
=(\psi_{k1\sigma},\ldots,\psi_{kN_S\sigma})^{\mathrm T}
$
are the Grassmann counterparts of the operators $\mathbf c_{\mathbf k\sigma}$ of
Eq.~\eqref{eq:H_momentum}, $\bar{\boldsymbol{\psi}}_{k\sigma}$ is the independent
conjugate field, and $k=(\mathbf k,\ii\omega_n)$ with $\omega_n=(2n+1)\pi T$.
In these variables the untwisted action corresponding to Eq.~\eqref{eq:H_momentum}
is
$
    \mathcal S=\sum_{k\sigma}\bar{\boldsymbol{\psi}}_{k\sigma}
    \left(-\ii\omega_n \mathbf{1} + \mathbf h_{\mathbf k\sigma}-\mu \mathbf{1}\right)
    \boldsymbol{\psi}_{k\sigma}
    -\frac{U}{\beta N_c}\sum_{QS}\bar b_S(Q) b_S(Q),
$
with
$
\bar b_S(Q)=\sum_k\bar\psi_{k+Q,S\uparrow}\bar\psi_{-k,S\downarrow}
$
the Grassmann counterpart of $b_{S\mathbf Q}^\dagger$, and
$Q=(\mathbf Q,\ii\nu_m)$ with $\nu_m=2\pi mT$.
The quartic term is decoupled by a Hubbard-Stratonovich transformation in the
pairing channel, channel by channel and momentum by momentum.
Introducing the pairing fields $\Phi_S(Q)$ through
$
\Phi_{lS}(\tau)=\frac{1}{\sqrt{\beta N_c}}\sum_Q
e^{\ii(\mathbf Q\cdot\mathbf r_{lS}-\nu_m\tau)}\Phi_S(Q),
$
which couple linearly to $\bar b_S(Q)$, gives
\begin{align}
    \mathcal S_{\mathrm{HS}}
    &=\sum_{k\sigma}\bar{\boldsymbol{\psi}}_{k\sigma}
    \left(-\ii\omega_n \mathbf{1} +\mathbf h_{\mathbf k\sigma}-\mu \mathbf{1} \right)
    \boldsymbol{\psi}_{k\sigma}
    +\sum_{QS}\frac{|\Phi_S(Q)|^2}{U}
    \nonumber \\
    &+\frac{1}{\sqrt{\beta N_c}}\sum_{kQS}
    \left[
    \Phi_S(Q)\bar\psi_{k+Q,S\uparrow}\bar\psi_{-k,S\downarrow}
    +\mathrm{c.c.}
    \right],
\label{eq:SHS}
\end{align}
with
$
\mathcal Z=\int\mathcal D[\bar{\boldsymbol{\psi}},\boldsymbol{\psi},
\Phi,\Phi^*] e^{-\mathcal S_{\mathrm{HS}}}$, where $\Phi\equiv\{\Phi_S(Q)\}
$
denotes the collection of pairing fields. Gaussian integration over
$\Phi$ returns the quartic action above, which fixes the relative sign
of the last two terms of Eq.~\eqref{eq:SHS}; with this choice a
uniform static field
$
\Phi_S(Q)=\sqrt{\beta N_c} \Delta_S\delta_{Q0}
$
reproduces Eq.~\eqref{eq:pairing_term}, where $\delta_{Q0}$ is a Kronecker delta.
For any static field, $\Phi_S(Q)\propto\delta_{\nu_m,0}$, the $\tau$ dependence
of $\Phi_{lS}(\tau)$ drops out, and we then write simply $\Phi_{lS}$; in
particular the uniform static field above gives $\Phi_{lS}=\Delta_S$, and the
$\sqrt{\beta N_c}$ in the Fourier convention is what makes the real-space field
carry the normalization of Eq.~\eqref{eq:Delta_def}.

Equation~\eqref{eq:SHS} is quadratic in the fermions, which can therefore be
integrated out exactly. Collecting them into the Nambu spinor
$
\mathbf C_k=(\boldsymbol{\psi}_{k\uparrow},
\bar{\boldsymbol{\psi}}_{-k\downarrow})^{\mathrm T},
$
the Grassmann counterpart
of the spinor of Sec.~\ref{sec:thermo} at zero twist, the fermionic part of
Eq.~\eqref{eq:SHS} reads
$
-\sum_{kk'}\bar{\mathbf C}_k\mathbf G^{-1}[\Phi]_{kk'}\mathbf C_{k'},
$
with
\begin{equation}
    \mathbf G^{-1}[\Phi]_{kk'}
    =\delta_{kk'}\big[\ii\omega_n \mathbf{1} -\mathbf h_{\mathrm{BdG}}^{(N)}(\mathbf k)\big]
    -\sum_{\alpha=1}^{2N_S}
    \frac{\Phi_\alpha(k-k')}{\sqrt{\beta N_c}} \mathbf V_\alpha
    \label{eq:Ginv}
\end{equation}
the inverse Nambu propagator, where $\mathbf h_{\mathrm{BdG}}^{(N)}$ is the
normal-state BdG matrix, i.e.\ Eq.~\eqref{eq:HMF_HBdG} at
$\boldsymbol{\Delta}=\mathbf 0$ and $\mathbf q=\mathbf 0$, the doubled
pairing field is
$
\boldsymbol{\Phi}(Q)=\big(\Phi_1(Q),\ldots,\Phi_{N_S}(Q),
\Phi_1^*(-Q),\ldots,\Phi_{N_S}^*(-Q)\big)^{\mathrm T}
$
with components $\Phi_\alpha(Q)$, $\alpha=1,\ldots,2N_S$,
and the pairing vertices $\mathbf V_\alpha$ are defined in
Appendix~\ref{app:MH}.
A pairing field of four-momentum $Q$ thus connects two fermionic
propagators whose momenta and frequencies differ by $Q$. Performing the
Grassmann integration gives the exact bosonic action
\begin{equation}
    \mathcal S_{\mathrm{eff}}[\Phi]
    =\frac{1}{U}\sum_{Q S}|\Phi_S(Q)|^2
    -\operatorname{Tr}\ln\left[-\mathbf G^{-1}[\Phi]\right],
    \label{eq:Seff}
\end{equation}
with
$
\mathcal Z=\int\mathcal D[\Phi,\Phi^*] e^{-\mathcal S_{\mathrm{eff}}[\Phi]}
$
and $\operatorname{Tr}$ the functional trace over momentum, Matsubara frequency
and Nambu-orbital indices; this is the only meaning of $\operatorname{Tr}$ in
what follows.

\subsection{Static and uniform Gaussian fluctuations}
\label{sec:below_Tc}

Below $T_c$, we expand about the uniform self-consistent saddle as
\begin{equation}
    \Phi_S(Q)=\sqrt{\beta N_c} \Delta_{0S}\delta_{Q0}+\eta_S(Q),
    \label{eq:saddle_expansion}
\end{equation}
and collect the fluctuations
$
\boldsymbol{\eta}(Q)=(\eta_1(Q),\ldots,\eta_{N_S}(Q))^{\mathrm T}
$
into the doubled field
$
\delta\boldsymbol{\Phi}(Q)=\big(\boldsymbol{\eta}(Q),
\boldsymbol{\eta}^*(-Q)\big)^{\mathrm T},
$
so that the doubled field of Sec.~\ref{sec:effaction} is
$
\boldsymbol{\Phi}(Q)=\sqrt{\beta N_c} \boldsymbol{\Psi}_0\delta_{Q0}
+\delta\boldsymbol{\Phi}(Q).
$
The two sectors of $\delta\boldsymbol{\Phi}(Q)$ are not independent but
related by
\begin{equation}
    \delta\boldsymbol{\Phi}(-Q)
    =\boldsymbol{\tau}_1\delta\boldsymbol{\Phi}^*(Q),
    \qquad
    \boldsymbol{\tau}_1=
    \begin{pmatrix}
        0 & \mathbf{1}\\
        \mathbf{1} & 0
    \end{pmatrix},
    \label{eq:doubling_constraint}
\end{equation}
where $\boldsymbol{\tau}_1$ exchanges the two sectors of the doubled pairing
space. Expanding the effective action about the saddle gives
$
\mathcal S_{\mathrm{eff}}=\mathcal S_{\mathrm{eff}}^{(0)}+\mathcal S_{\mathrm{eff}}^{(2)}+\cdots,
$
where
$
\mathcal S_{\mathrm{eff}}^{(0)}=\beta N_c
\Omega_0(\mathbf 0,\boldsymbol{\Psi}_0)
$
is the mean-field value and the linear term vanishes by the saddle-point
condition. Since the saddle is translationally invariant, the quadratic
action is diagonal in four-momentum,
\begin{equation}
    \mathcal S_{\mathrm{eff}}^{(2)}
    =\frac{1}{2}\sum_Q\delta\boldsymbol{\Phi}^\dagger(Q)
    \mathbf M(Q)\delta\boldsymbol{\Phi}(Q),
    \label{eq:S2}
\end{equation}
which defines the Gaussian pairing-fluctuation kernel $\mathbf M(Q)$, equivalently
the inverse fluctuation propagator.

Equation~\eqref{eq:S2} contains the thermodynamic Hessian of
Sec.~\ref{sec:thermo} as its zero-four-momentum limit. For a static and
spatially uniform fluctuation, $Q=(\mathbf{0},0)$, the Fourier convention gives
$
\delta\boldsymbol{\Phi}(\mathbf 0,0)=\sqrt{\beta N_c} \delta\boldsymbol{\Psi}
$
with
$\delta\boldsymbol{\Psi}$ the uniform variation of $\boldsymbol{\Psi}_0$,
so that only the single term $Q = (\mathbf{0}, 0)$ survives in
Eq.~\eqref{eq:S2} and
$
\mathcal S_{\mathrm{eff}}^{(2)}=\frac{\beta N_c}{2}\delta\boldsymbol{\Psi}^\dagger
\mathbf M(\mathbf 0,0)\delta\boldsymbol{\Psi}.
$
On the other hand, for static
and spatially uniform configurations the effective action is the mean-field
potential itself,
$
\mathcal S_{\mathrm{eff}}=\beta N_c\Omega_0(\mathbf 0,\boldsymbol{\Psi}),
$
whose expansion~\eqref{eq:expansion} gives
$
\mathcal S_{\mathrm{eff}}^{(2)}=\frac{\beta N_c}{2}\delta\boldsymbol{\Psi}^\dagger
\boldsymbol{\mathcal H}(\mathbf 0)\delta\boldsymbol{\Psi}.
$
Comparing the two quadratic forms yields
\begin{equation}
    \mathbf M(\mathbf 0,0)=\boldsymbol{\mathcal{H}}(\mathbf 0).
    \label{eq:M_and_H}
\end{equation}
The thermodynamic Hessian is therefore not merely analogous to the fluctuation
kernel: it is that kernel at zero momentum and zero frequency.
At finite momentum or frequency $\mathbf M(Q)$ carries spatial and dynamical
information that is absent from $\boldsymbol{\mathcal H}(\mathbf 0)$, so the
identification is static. A microscopic derivation of Eq.~\eqref{eq:M_and_H}
from the effective action is given in Appendix~\ref{app:MH}.

We now restrict to static pairing fluctuations and define
$
\mathbf M(\mathbf Q) \equiv \mathbf M(\mathbf Q,\ii\nu_m=0).
$
Starting from the microscopic representation of $\mathbf M(Q)$ derived in
Appendix~\ref{app:MH}, performing the Matsubara sum and evaluating the
trace in the eigenbasis of $\mathbf h_{\mathrm{BdG}}(\mathbf k,\mathbf 0)$
gives the static kernel in a form that can be evaluated directly,
\begin{align}
    &  [M(\mathbf Q)]_{\alpha\alpha'}
    =\frac{\delta_{\alpha\alpha'}}{U}
    +\frac{1}{N_c}\sum_{\mathbf k}\sum_{pp'}
    \langle p',\mathbf k|\mathbf V_\alpha^\dagger|p,\mathbf k{+}\mathbf Q\rangle
    \nonumber \\
    &\times
    \langle p,\mathbf k{+}\mathbf Q|\mathbf V_{\alpha'}|p',\mathbf k\rangle
    \frac{f[E_p(\mathbf k{+}\mathbf Q)]-f[E_{p'}(\mathbf k)]}
    {E_p(\mathbf k{+}\mathbf Q)-E_{p'}(\mathbf k)},
\label{eq:MQ_explicit}
\end{align}
where $|p,\mathbf k\rangle$ and $E_p(\mathbf k)\equiv E_p(\mathbf k,\mathbf 0)$
are the eigenvectors and eigenvalues of the untwisted BdG matrix
$\mathbf h_{\mathrm{BdG}}(\mathbf k,\mathbf 0)$, $f$ is the Fermi
function, and the ratio is replaced by its derivative $f'(E_p)$ for coincident
energies. Since $\mathbf h_{\mathbf k+\mathbf Q,\sigma}$ is known analytically,
Eq.~\eqref{eq:MQ_explicit} can be evaluated at any real $\mathbf Q$, which need
not be commensurate with the $\mathbf k$ grid.

The static kernel is Hermitian for real $\mathbf Q$, and the
constraint~\eqref{eq:doubling_constraint} on the doubled field implies the
particle-hole relation
\begin{equation}
    \boldsymbol{\tau}_1\mathbf M^*(\mathbf Q)\boldsymbol{\tau}_1
    =\mathbf M(-\mathbf Q).
    \label{eq:ph_kernel}
\end{equation}
Differentiating Eq.~\eqref{eq:ph_kernel} at $\mathbf Q=\mathbf 0$ gives
$
\boldsymbol{\tau}_1(\partial_i\mathbf M)^*\boldsymbol{\tau}_1
=-\partial_i\mathbf M
$
and
$
\boldsymbol{\tau}_1(\partial_i\partial_j\mathbf M)^*
\boldsymbol{\tau}_1=+\partial_i\partial_j\mathbf M,
$
where
$
\partial_i\equiv \partial/\partial Q_i|_{\mathbf Q=\mathbf 0}
$
denotes the pair-momentum derivative at zero pair momentum. Throughout,
a bare $\partial_i$ carries this meaning; derivatives at general
$\mathbf Q$ are written $\partial_{Q_i}$, derivatives with respect to the
fermionic twist $\partial_{q_i}$, and real-space gradients $\nabla_i$.
By Eq.~\eqref{eq:M_and_H}, the zero mode $\mathbf u_0$ of Eq.~\eqref{eq:u0} is
also a zero mode of the static kernel,
$\mathbf M(\mathbf 0)\mathbf u_0=\mathbf 0$, at every temperature at which the
symmetry is broken. We normalize it as
$
\mathbf w_0=\mathbf u_0/(\sqrt2|\boldsymbol{\Delta}_0|)
$
and complete the eigenbasis with orthonormal massive modes satisfying
\begin{equation}
    \mathbf M(\mathbf 0)\mathbf w_n=\lambda_n\mathbf w_n,
    \qquad n>0,
    \label{eq:massive_modes}
\end{equation}
with
$
\langle\mathbf w_m|\mathbf w_n\rangle=\delta_{mn}.
$
The $\lambda_n$ are the
static stiffnesses of the amplitude and relative-phase channels;
the latter correspond to Leggett modes~\cite{leggett66, Blumberg07}
when their dynamical continuation is well defined, although no statement about
collective-mode dynamics is implied here.
At a stable saddle $\lambda_n>0$, and in the following we assume that the
$U(1)$ Goldstone mode is the only zero mode of $\mathbf M(\mathbf 0)$; at a
secondary instability one of the massive modes may itself soften.

\subsection{Goldstone projection and low-$\mathbf{Q}$ expansion}
\label{sec:lowQ}

The eigenvectors $\{\mathbf w_0,\mathbf w_n\}$ with ${n>0}$ form a complete
basis in the doubled pairing space. We use this fixed zero-momentum basis
to describe a slowly varying static fluctuation, writing
$
\delta\boldsymbol{\Phi}(\mathbf Q)\equiv
\delta\boldsymbol{\Phi}(\mathbf Q,\ii\nu_m=0)
$
and decomposing
\begin{equation}
    \delta\boldsymbol{\Phi}(\mathbf Q)=\theta(\mathbf Q)\mathbf w_0
    +\sum_{n>0}\chi_n(\mathbf Q)\mathbf w_n ,
    \label{eq:decomposition}
\end{equation}
where $\theta(\mathbf Q)$ is the Goldstone-mode amplitude and
$\boldsymbol{\chi}(\mathbf Q)$ collects the massive ones. In this basis the
static kernel becomes a block matrix with elements
$
M_{00}(\mathbf Q)
=\langle\mathbf w_0|\mathbf M(\mathbf Q)|\mathbf w_0\rangle,
$
$
[M_{\perp 0}(\mathbf Q)]_n
=\langle\mathbf w_n|\mathbf M(\mathbf Q)|\mathbf w_0\rangle,
$
$
[M_{0\perp}(\mathbf Q)]_n
=\langle\mathbf w_0|\mathbf M(\mathbf Q)|\mathbf w_n\rangle,
$
and
$
[M_{\perp\perp}(\mathbf Q)]_{mn}
=\langle\mathbf w_m|\mathbf M(\mathbf Q)|\mathbf w_n\rangle.
$
Collecting the amplitudes into
$
\boldsymbol{\zeta}(\mathbf Q)=\big(\theta(\mathbf Q),
\boldsymbol{\chi}(\mathbf Q)\big)^{\mathrm T},
$
the quadratic action reads
\begin{equation}
    \mathcal S_{\mathrm{eff}}^{(2)}=\frac{1}{2}\sum_{\mathbf Q}
    \boldsymbol{\zeta}^\dagger(\mathbf Q)
    \begin{bmatrix}
        M_{00}(\mathbf Q)
        & \mathbf M_{0\perp}(\mathbf Q) \\
        \mathbf M_{\perp 0}(\mathbf Q)
        & \mathbf M_{\perp\perp}(\mathbf Q)
    \end{bmatrix}
    \boldsymbol{\zeta}(\mathbf Q).
    \label{eq:S2_blocks}
\end{equation}
The Goldstone mode is coupled to the massive modes through the off-diagonal
blocks, which vanish at $\mathbf Q=\mathbf 0$ because
$\{\mathbf w_0,\mathbf w_n\}$ diagonalize $\mathbf M(\mathbf 0)$. Since
$\mathbf M_{\perp\perp}(\mathbf 0)$ has eigenvalues $\lambda_n>0$,
that block remains invertible for sufficiently small $\mathbf Q$, and the massive
amplitudes can be integrated out. At quadratic order this is equivalent to
eliminating them through their stationary values,
$
\boldsymbol{\chi}(\mathbf Q)
=-\mathbf M_{\perp\perp}^{-1}(\mathbf Q)
\mathbf M_{\perp 0}(\mathbf Q)\theta(\mathbf Q),
$
which leaves
$
\mathcal S_{\mathrm{eff}}^{(2)}=\frac{1}{2}\sum_{\mathbf Q}\theta^*(\mathbf Q)
M_{\mathrm{eff}}(\mathbf Q)\theta(\mathbf Q)
$
with
\begin{equation}
    M_{\mathrm{eff}}(\mathbf Q)=M_{00}(\mathbf Q)
    -\mathbf M_{0\perp}(\mathbf Q)
    \mathbf M_{\perp\perp}^{-1}(\mathbf Q)
    \mathbf M_{\perp 0}(\mathbf Q).
    \label{eq:Meff}
\end{equation}
Thus $M_{\mathrm{eff}}(\mathbf Q)$ is the Schur complement of the
massive sector, and represents the energy cost of a slowly varying
Goldstone-mode fluctuation after all massive pairing degrees of freedom have
relaxed. Since $\mathbf M_{\perp\perp}(\mathbf Q)$ is positive
definite in a stable phase, so is its inverse, and the subtracted term
in Eq.~\eqref{eq:Meff} is non-negative: the relaxation can only lower the
Goldstone stiffness relative to its frozen-pairing value, in the same way
that two springs in series are softer than either spring alone.

We now expand $M_{\mathrm{eff}}(\mathbf Q)$ about $\mathbf Q=\mathbf 0$. The
zeroth-order term vanishes because $\mathbf w_0$ is the Goldstone mode. Since
$\mathbf w_0$ is invariant under $\boldsymbol{\tau}_1$ complex conjugation,
Eq.~\eqref{eq:ph_kernel} implies that $M_{00}(\mathbf Q)$ is real and
even in $\mathbf Q$, so that
$
\langle\mathbf w_0|\partial_i\mathbf M|\mathbf w_0\rangle=0
$
without invoking inversion or time-reversal symmetry. Each mixed block is of
order $\mathbf{Q}$, so that at order $\mathbf{Q}^2$ only
$
[M_{\perp\perp}^{-1}(\mathbf 0)]_{mn} =\delta_{mn}/\lambda_n
$
is needed and derivatives of
$\mathbf M_{\perp\perp}^{-1}$ do not enter.
Equation~\eqref{eq:Meff} then gives
$
M_{\mathrm{eff}}(\mathbf Q)
=\sum_{ij}Q_iQ_j \mathcal Q_{ij}+O(\mathbf{Q}^4),
$
with
\begin{align}
\mathcal Q_{ij}=\frac{1}{2}
&\langle\mathbf w_0|
\partial_i\partial_j\mathbf M|\mathbf w_0\rangle
\nonumber \\
&-
\sum_{n>0}\frac{\operatorname{Re} \left[
\langle\mathbf w_0|\partial_i\mathbf M|\mathbf w_n\rangle
\langle\mathbf w_n|\partial_j\mathbf M|\mathbf w_0\rangle
\right]}
{\lambda_n}.
\label{eq:Qij}
\end{align}
Here $\operatorname{Re}$ denotes the real part, and $\mathcal Q_{ij}$ is real and
symmetric. The first term is the direct curvature of the Goldstone sector, and
the second accounts for its coupling to, and relaxation through, the massive
pairing modes. If the uniform saddle is locally stable against static
fluctuations at small finite momentum, then
$\mathbf M(\mathbf Q)\succeq0$ near $\mathbf Q=\mathbf 0$ and
consequently $\boldsymbol{\mathcal Q}\succeq0$; a negative eigenvalue of
$\boldsymbol{\mathcal Q}$ signals an instability toward finite pair momentum.

Equation~\eqref{eq:Qij} is nothing but nondegenerate
Rayleigh-Schr\"odinger perturbation theory for the branch $\lambda_0(\mathbf Q)$
of $\mathbf M(\mathbf Q)$ that evolves continuously from the
Goldstone mode, with $\lambda_0(\mathbf 0)=0$: treating
$\mathbf M(\mathbf 0)$ as the unperturbed operator and
$\mathbf M(\mathbf Q)-\mathbf M(\mathbf 0)$, expanded to second order in
$\mathbf Q$, as the perturbation, the first- and second-order shifts of that
level, with energy denominators
$
\lambda_0(\mathbf 0)-\lambda_n=-\lambda_n,
$
reproduce the two terms of Eq.~\eqref{eq:Qij} once symmetrized in $i$ and $j$.
Since $\lambda_0(\mathbf Q)$ is even in $\mathbf Q$ by
Eq.~\eqref{eq:ph_kernel}, the omitted terms are of order $\mathbf{Q}^4$, and therefore
\begin{equation}
    \mathcal Q_{ij}
    =
    \frac{1}{2}
    \frac{\partial^2\lambda_0(\mathbf Q)}
    {\partial Q_i\partial Q_j}
    \bigg|_{\mathbf Q=\mathbf 0},
    \label{eq:Qij_soft_curvature}
\end{equation}
which is a basis-independent definition of $\mathcal Q_{ij}$: an eigenvalue is
invariant under any unitary change of the channel basis, including a
$\mathbf Q$-dependent one, whereas the two terms of Eq.~\eqref{eq:Qij}
separately are not. The Schur complement along the fixed direction $\mathbf w_0$
and the exact eigenvalue differ only at order $\mathbf{Q}^4$.

\subsection{Quantum geometry of Goldstone modes}
\label{sec:pairgeometry}

We next identify the geometric origin of the relaxation term. The eigenvector
$\mathbf w_0(\mathbf Q)$ of the soft branch satisfies
$
\mathbf M(\mathbf Q)\mathbf w_0(\mathbf Q)
=\lambda_0(\mathbf Q)\mathbf w_0(\mathbf Q),
$
with
$\mathbf w_0(\mathbf 0)=\mathbf w_0$. Differentiating at $\mathbf Q=\mathbf 0$,
using $\lambda_0(\mathbf 0)=0$ and
$
\partial_i\lambda_0=\langle\mathbf w_0|\partial_i\mathbf M
|\mathbf w_0\rangle=0,
$
and projecting onto a massive eigenmode gives
\begin{equation}
    \langle\mathbf w_n|\partial_i\mathbf M|\mathbf w_0\rangle
    =-\lambda_n\langle\mathbf w_n|\partial_i\mathbf w_0\rangle,
    \qquad n>0 .
    \label{eq:eigvec_derivative}
\end{equation}
Inserting this into Eq.~\eqref{eq:Qij} converts the second-order eigenvalue
shift into a stiffness-weighted overlap of eigenvector derivatives. As the pair
momentum is varied, the soft eigenvector $\mathbf w_0(\mathbf Q)$ traces a curve
in the doubled pairing space. Resolving its tangent along the massive branches
$\mathbf w_n(\mathbf Q)$, i.e.\ the eigenvectors of
$\mathbf M(\mathbf Q)$ that evolve from $\mathbf w_n$, defines the
mode-resolved pair quantum metric of the Goldstone mode,
\begin{equation}
    g_{0\mathbf{Q}}^{n ij}=2\operatorname{Re}\left[
    \langle\partial_{Q_i}\mathbf w_0(\mathbf Q)|\mathbf w_n(\mathbf Q)\rangle
    \langle\mathbf w_n(\mathbf Q)|\partial_{Q_j}\mathbf w_0(\mathbf Q)\rangle
    \right].
    \label{eq:g0n}
\end{equation}
Equation~\eqref{eq:g0n}
is a metric on the manifold of pairing configurations swept out by $\mathbf Q$
rather than on the Bloch manifold swept out by $\mathbf k$. Only its value at
zero pair momentum enters below, and with it Eq.~\eqref{eq:Qij} becomes
\begin{equation}
    \mathcal Q_{ij}=\frac{1}{2}
    \langle\mathbf w_0|\partial_i\partial_j\mathbf M|\mathbf w_0\rangle
    -\frac{1}{2} \sum_{n>0}\lambda_n g_{0\mathbf{0}}^{n ij}.
    \label{eq:Q_metric}
\end{equation}
The relaxation correction is therefore the stiffness-weighted pair quantum
metric $\sum_{n>0}\lambda_n g_{0\mathbf{0}}^{n ij}$.
It is gauge invariant under separate rephasings of $\mathbf w_0$ and
$\mathbf w_n$, because $\langle\mathbf w_0|\mathbf w_n\rangle=0$, and it requires
the Goldstone branch to remain nondegenerate near $\mathbf Q=\mathbf 0$, so
that $\mathbf w_0(\mathbf Q)$ can be continued smoothly. Three geometric objects
should be distinguished here. The Bloch quantum metric of the band structure
enters only indirectly, through the interband matrix elements of
$\partial_i\partial_j\mathbf M$~\cite{huhtinen22}. The unweighted sum
$
\sum_{n>0}g_{0\mathbf{0}}^{n ij}
=2\operatorname{Re}\left[\langle\partial_i\mathbf w_0|
(\mathbf{1}-|\mathbf w_0\rangle\langle\mathbf w_0|)|\partial_j\mathbf w_0\rangle\right]
$
is the full Fubini-Study metric of the soft eigenvector, a property of that
eigenvector alone. What enters the superfluid weight is neither of these, but
the sum in which each massive channel is weighted by its own static stiffness
$\lambda_n$
~\footnote{
The same mixing defines a mode-resolved pair Berry curvature,
$
\mathcal F_{0\mathbf Q}^{n ij}=-2\operatorname{Im}\left[
\langle\partial_{Q_i}\mathbf w_0(\mathbf Q)|\mathbf w_n(\mathbf Q)\rangle
\langle\mathbf w_n(\mathbf Q)|\partial_{Q_j}\mathbf w_0(\mathbf Q)\rangle
\right],
$
the antisymmetric counterpart of Eq.~\eqref{eq:g0n}. Since
Eq.~\eqref{eq:ph_kernel} relates $\mathbf w_n(-\mathbf Q)$ to
$\boldsymbol\tau_1\mathbf w_n^*(\mathbf Q)$, the curvature is odd in
$\mathbf Q$ while the metric is even; hence
$\mathcal F_{0\mathbf 0}^{n ij}=0$, and the pair geometry of the Goldstone
mode at zero pair momentum is entirely metric. The same oddness forces the
Chern number of every isolated branch of $\mathbf M(\mathbf Q)$ over the BZ
of the pair momentum to vanish below $T_c$, whether or not time-reversal
symmetry is present.
}.
\section{Superfluid weight}
\label{sec:supfwe}

Two facts connect the thermodynamic formulation of Sec.~\ref{sec:thermo} to the
fluctuation kernel of Sec.~\ref{sec:effective_action}. The first is the
field-theoretical reading of the twist. The twisted mean-field problem of
Sec.~\ref{sec:thermo} is not a new theory: it is the untwisted exact
action~\eqref{eq:Seff} evaluated on the particular configuration in which the
pairing field is a single static Fourier mode of momentum $2\mathbf q$, i.e.\ in
which it winds in real space,
\begin{equation}
\begin{gathered}
    \Phi_S(Q)=\sqrt{\beta N_c} \Delta_S
    \delta_{\mathbf Q,2\mathbf q} \delta_{\nu_m,0},
    \quad
    \Phi_{lS}=\Delta_S e^{2\ii\mathbf q\cdot\mathbf r_{lS}},
    \\
    \boldsymbol{\Phi}(\mathbf r)=\mathbf U_{\mathbf q}(\mathbf r)
    \boldsymbol{\Psi},
    \quad
    \mathbf U_{\mathbf q}(\mathbf r)
    =e^{2\ii\mathbf q\cdot\mathbf r \boldsymbol{\tau}_3},
    \quad
    \boldsymbol{\tau}_3=\operatorname{diag}(\mathbf{1},-\mathbf{1}),
\end{gathered}
\label{eq:winding}
\end{equation}
where $\boldsymbol{\Phi}(\mathbf r)$ denotes the doubled real-space pairing
field at position $\mathbf r=\mathbf r_{lS}$, the conjugate component carrying
the opposite phase, and
$\boldsymbol{\Psi}=(\boldsymbol{\Delta},\boldsymbol{\Delta}^*)^{\mathrm T}$
is an arbitrary uniform pairing vector. By definition of the saddle-point
action,
\begin{equation}
    \mathcal S_{\mathrm{eff}}[\mathbf U_{\mathbf q}\boldsymbol{\Psi}]
    =\beta N_c \Omega_0(\mathbf q,\boldsymbol{\Psi})
    \label{eq:gauge}
\end{equation}
for every $\boldsymbol{\Psi}$, which at $\mathbf q=\mathbf 0$ and
$\boldsymbol{\Psi}=\boldsymbol{\Psi}_0$ reduces to
$\mathcal S_{\mathrm{eff}}^{(0)}$ of Sec.~\ref{sec:below_Tc}, since
$\mathbf U_{\mathbf 0}=\mathbf{1}$. The gauge
transformation of Sec.~\ref{sec:thermo}, which moved the phase from the pairing
field onto the fermions, is therefore the statement that the fermionic
determinant of this configuration is that of
$\mathbf h_{\mathrm{BdG}}(\mathbf k,\mathbf q)$. Every $\mathbf q$ derivative of
$\Omega_0(\mathbf q,\cdot)$ is consequently a functional derivative of the
untwisted action along the winding configuration, and such derivatives at
$\boldsymbol{\Psi}_0$ are given by the kernel of Sec.~\ref{sec:below_Tc}. The
second fact is the identification~\eqref{eq:M_and_H} of the thermodynamic
Hessian with that kernel at zero momentum and zero frequency. With these two
facts both terms of Eq.~\eqref{eq:Ds_thermo} can be evaluated.

Using Eq.~\eqref{eq:winding} together with translational invariance, as detailed
in Appendix~\ref{app:Bi}, the mixed derivative introduced in
Sec.~\ref{sec:thermo} becomes
\begin{equation}
    \mathbf B_i=-2\ii(\partial_i\mathbf M)\mathbf u_0 .
    \label{eq:B_i}
\end{equation}
Projecting Eq.~\eqref{eq:hfc} onto a massive eigenmode $\mathbf w_n$ and using
Eqs.~\eqref{eq:B_i} and \eqref{eq:eigvec_derivative} gives, for every $n>0$,
\begin{equation}
    \lambda_n\langle\mathbf w_n|\mathbf F_i\rangle
    =-2\sqrt2 \ii|\boldsymbol{\Delta}_0|\;
    \lambda_n\langle\mathbf w_n|\partial_i\mathbf w_0\rangle .
    \label{eq:F_projected}
\end{equation}
At a stable saddle $\lambda_n>0$, so the stiffness cancels on both sides and the
response is fixed up to its undetermined Goldstone component,
\begin{equation}
    \mathbf P_\perp\mathbf F_i
    =-2\sqrt2 \ii|\boldsymbol{\Delta}_0|\;
    \mathbf P_\perp|\partial_i\mathbf w_0\rangle,
    \quad
    \mathbf P_\perp=\mathbf{1}-|\mathbf w_0\rangle\langle\mathbf w_0| .
    \label{eq:F_rotation}
\end{equation}
That is, the self-consistent response of the pairing vector to the twist is the
transverse part of the pair-momentum derivative of the soft eigenvector: as
$\mathbf Q=2\mathbf q$ is turned on, the gaps follow the rotation of
$\mathbf w_0(\mathbf Q)$. Both sides of Eq.~\eqref{eq:F_rotation} carry the same
arbitrariness: $\mathbf F_i$ is undetermined along $\mathbf u_0$ because of
global $U(1)$, and $\mathbf w_0(\mathbf Q)$ is defined only up to a
$\mathbf Q$-dependent phase. The projector $\mathbf P_\perp$ removes both, so
the identification is between gauge-covariant objects.
Contracting Eq.~\eqref{eq:F_rotation} with $\boldsymbol{\mathcal H}(\mathbf 0)$,
which annihilates $\mathbf u_0$, gives the relaxation term in geometric form,
\begin{equation}
    -\mathbf F_i^\dagger\boldsymbol{\mathcal H}(\mathbf 0)\mathbf F_j
    =-4|\boldsymbol{\Delta}_0|^2\sum_{n>0}\lambda_n g_{0\mathbf{0}}^{n ij},
    \label{eq:relaxation_metric}
\end{equation}
the stiffness-weighted pair quantum metric of Eq.~\eqref{eq:Q_metric}. This
provides a geometric interpretation of the relaxation correction of
Ref.~\cite{huhtinen22}.

The same gauge relation connects the frozen-pairing curvature to
the direct Goldstone-sector curvature,
\begin{equation}
    V[D_{\mathrm{frozen}}]_{ij}
    =4N_c|\boldsymbol{\Delta}_0|^2
    \langle\mathbf w_0|\partial_i\partial_j\mathbf M|\mathbf w_0\rangle ,
    \label{eq:frozen_kernel}
\end{equation}
as shown in Appendix~\ref{app:Bi}. Combining Eqs.~\eqref{eq:frozen_kernel} and
\eqref{eq:relaxation_metric} with Eq.~\eqref{eq:Ds_thermo}, and using
Eq.~\eqref{eq:Q_metric}, gives
\begin{equation}
    V[D_s]_{ij}=8N_c|\boldsymbol{\Delta}_0|^2 \mathcal Q_{ij} .
    \label{eq:Ds_Qij}
\end{equation}
The prefactor $8$ is the product of the Nambu norm
$\langle\mathbf u_0|\mathbf u_0\rangle/|\boldsymbol{\Delta}_0|^2=2$ of the
Goldstone vector and a factor $4$ from $\nabla_i\phi=2q_i$. For the pyrochlore model of
Ref.~\cite{iskin24}, where the relaxation term vanishes by symmetry (see
below), Eq.~\eqref{eq:Ds_Qij} reduces at $T=0$ to the Goldstone-dispersion
relation obtained there, including this factor of $8$. The present formulation
extends that relation to arbitrary $T<T_c$ and to sublattice-dependent pairing,
for which the massive channels need not decouple and $\mathcal Q_{ij}$ carries
the pair-geometric term of Eq.~\eqref{eq:Q_metric}.
The amplitude appearing in Eq.~\eqref{eq:Ds_Qij} is the squared pairing field in
the normalization of Eq.~\eqref{eq:Delta_def}; since $\mathcal Q_{ij}$ scales
inversely under a rescaling of that field, only the product
$|\boldsymbol{\Delta}_0|^2\mathcal Q_{ij}$ is independent of the choice.

The relaxation term vanishes precisely when
$
\langle\mathbf w_n|\partial_i\mathbf M|\mathbf w_0\rangle=0
$
for all $n>0$, equivalently $\mathbf P_\perp\partial_i\mathbf w_0=\mathbf 0$,
in which case Eq.~\eqref{eq:Ds_Qij} reduces to the frozen-pairing
expression~\eqref{eq:frozen_kernel}, the uniform-pairing result of
Refs.~\cite{peotta15, liang17}. This is a condition on the twisted problem,
and it is not implied by equal gap magnitudes at $\mathbf q=\mathbf 0$.
Nor is it implied by time reversal, which maps
$\boldsymbol{\Delta}_0(\mathbf q)$ onto $\boldsymbol{\Delta}_0^*(-\mathbf q)$
and therefore only requires $\mathbf F_i$ to be purely imaginary in a gauge
where $\boldsymbol{\Delta}_0$ is real, i.e. the relative phases of the gaps
to develop linearly in $\mathbf q$; nor by an inversion that permutes the
orbitals. It does follow from an inversion that maps every orbital onto
itself, provided the orbitals are placed at their inversion-symmetric
positions in the Fourier convention of Sec.~\ref{sec:thermo}:
$\boldsymbol{\Delta}_0(\mathbf q)$ and $\boldsymbol{\Delta}_0(-\mathbf q)$
then coincide up to a global phase, so that $\mathbf F_i$ is purely
longitudinal and $\mathbf P_\perp\mathbf F_i=\mathbf 0$. Like the two terms
of Eq.~\eqref{eq:Q_metric} separately, the decoupling condition is thus a
property of a given orbital embedding (Appendix~\ref{app:embedding}). It
holds for the lattices of Refs.~\cite{iskin19, iskin24, elden26} and for the
unstaggered Lieb lattice of Sec.~\ref{sec:numerics} at $\kappa=\tfrac12$.
At the other extreme, as a massive channel softens, $\lambda_n
g_{0\mathbf{0}}^{n ij}$ grows as $\lambda_n^{-1}$ by
Eq.~\eqref{eq:eigvec_derivative}, signaling the breakdown of the single-mode
elimination rather than a physical divergence: that channel must then be kept
with the Goldstone direction in an enlarged soft subspace. This happens as
$T\to T_c^-$, treated in Sec.~\ref{sec:Tc}.

\section{Critical regime}
\label{sec:Tc}

As $T\to T_c^-$ the amplitude mode of the critical channel becomes degenerate
with the Goldstone direction at zero eigenvalue, so that the ordered-state
kernel has a two-dimensional soft space; for the stationary zero-momentum
critical point considered here the two do not mix at first order in
$\mathbf Q$. As shown below, $\mathcal Q_{ij}$ has a finite critical limit,
and what vanishes at the transition is the prefactor
$|\boldsymbol{\Delta}_0|^2$ of Eq.~\eqref{eq:Ds_Qij}. Since the
Goldstone-based construction has no counterpart above $T_c$, we work here
with the normal-state pair propagator, to which the same soft-mode
projection applies and yields the GL gradient coefficient directly.

\subsection{GL gradient coefficient}
\label{sec:GL}

At the normal-state saddle the global $U(1)$ symmetry is unbroken and the
fluctuation field coincides with the pairing field, $\eta_S(Q)=\Phi_S(Q)$. The
Gaussian fluctuation kernel then takes the block-diagonal form
\begin{equation}
    \mathbf M(Q)=
    \begin{bmatrix}
        \boldsymbol{\Gamma}^{-1}(Q) & 0\\
        0 & [\boldsymbol{\Gamma}^{-1}(-Q)]^{\mathrm T}
    \end{bmatrix},
    \label{eq:M_normal}
\end{equation}
where the $N_S \times N_S$ matrix $\boldsymbol{\Gamma}^{-1}(Q)$ is the
normal-state inverse pair propagator. The two blocks give identical contributions after
relabeling $Q\to-Q$, so the quadratic effective action becomes
$
\mathcal S_{\mathrm{eff}}^{(2)}=\sum_Q\boldsymbol{\eta}^\dagger(Q)
\boldsymbol{\Gamma}^{-1}(Q)\boldsymbol{\eta}(Q).
$
As with the fluctuation kernel, we denote the static part of the inverse
pair propagator by
$
\mathbf K(\mathbf Q)\equiv
\boldsymbol{\Gamma}^{-1}(\mathbf Q,\ii\nu_m=0),
$
which is Hermitian. We assume a zero-momentum transition
with a single critical pairing channel and write
\begin{equation}
    \mathbf K(\mathbf 0;T) \mathbf v_n(T)
    =\lambda_n^{(N)}(T) \mathbf v_n(T),
    \label{eq:normal_modes}
\end{equation}
with
$
\langle\mathbf v_m(T)|\mathbf v_n(T)\rangle=\delta_{mn}.
$
At $T_c$ the Thouless criterion gives $\lambda_0^{(N)}(T_c)=0$, while
$\lambda_n^{(N)}(T_c)>0$ for $n>0$. In the following we write
$
\mathbf v_n\equiv\mathbf v_n(T_c)
$
and evaluate $\lambda_n^{(N)}$ at $T_c$ unless the temperature argument is displayed;
the superscript distinguishes these $N_S$ normal-state eigenvalues from the
stiffnesses $\lambda_n$ of the doubled ordered-state kernel in
Eq.~\eqref{eq:massive_modes}. The critical eigenvector $\mathbf v_0$ determines
the relative orbital structure of the pairing field at the transition.

Near $T_c$ we decompose the pairing field into the critical mode and the
remaining noncritical modes,
$
\boldsymbol{\eta}(\mathbf Q)=\Theta(\mathbf Q)\mathbf v_0
+\sum_{n>0}\rho_n(\mathbf Q)\mathbf v_n,
$
in analogy with Eq.~\eqref{eq:decomposition} but in the undoubled channel
space; the critical-mode amplitude $\Theta(\mathbf Q)$ and the noncritical
amplitudes $\rho_n(\mathbf Q)$ are distinct from the Goldstone and massive
amplitudes $\theta(\mathbf Q)$ and $\chi_n(\mathbf Q)$ of the ordered state.
In this basis the static kernel takes the block form
$
\mathbf K(\mathbf Q)=\Big[\begin{smallmatrix}
K_{00}(\mathbf Q)&\mathbf K_{0\perp}(\mathbf Q)\\
\mathbf K_{\perp0}(\mathbf Q)&\mathbf K_{\perp\perp}(\mathbf Q)
\end{smallmatrix}\Big],
$
with blocks defined as in Eq.~\eqref{eq:S2_blocks} with
$\mathbf w_n\to\mathbf v_n$.
Since $\lambda_n^{(N)}>0$ for $n>0$, $\mathbf K_{\perp\perp}(\mathbf Q)$
is invertible at $\mathbf Q=\mathbf 0$ and remains so for sufficiently small
$\mathbf Q$, and the noncritical modes can be integrated out, giving the
effective kernel of the critical mode,
\begin{equation}
    K_\mathrm{eff}(\mathbf Q)=K_{00}(\mathbf Q)
    -\mathbf K_{0\perp}(\mathbf Q)
    \mathbf K_{\perp\perp}^{-1}(\mathbf Q)
    \mathbf K_{\perp0}(\mathbf Q),
    \label{eq:Gamma_eff}
\end{equation}
the normal-state counterpart of Eq.~\eqref{eq:Meff}.

We define
$
a(T)=K_\mathrm{eff}(\mathbf 0;T),
$
with $a(T_c)=0$ by the critical-mode condition. For a zero-momentum
critical point, $\mathbf Q=\mathbf 0$ is assumed to be a stationary point of
the critical branch, so that
$\partial_iK_\mathrm{eff}=0$.
Below $T_c$, Eq.~\eqref{eq:ph_kernel} and the doubled particle-hole
structure imply the stronger condition that $M_{00}(\mathbf Q)$ is even in
$\mathbf Q$. No such evenness follows automatically in the undoubled
normal-state problem, although it may be enforced, for example, by inversion
or time-reversal symmetry. To leading order in the GL expansion,
$
K_\mathrm{eff}(\mathbf Q;T)=a(T)
+\tfrac12\sum_{ij}c_{ij}Q_iQ_j+\cdots,
$
with $c_{ij}=\partial_i\partial_jK_\mathrm{eff}$. Expanding
Eq.~\eqref{eq:Gamma_eff} to second order in $\mathbf Q$ gives
\begin{align}
c_{ij}=&\langle \mathbf v_0|\partial_i\partial_j
\mathbf K|\mathbf v_0\rangle
\nonumber \\
&-2\sum_{n>0}\frac{\operatorname{Re} \left[
\langle \mathbf v_0|\partial_i\mathbf K|\mathbf v_n\rangle
\langle \mathbf v_n|\partial_j\mathbf K|\mathbf v_0\rangle\right]}
{\lambda_n^{(N)}}.
\label{eq:cij_raw}
\end{align}
Differentiating the eigenvalue equation
of the critical branch $\mathbf v_0(\mathbf Q)$ of
$\mathbf K(\mathbf Q)$ at $\mathbf Q=\mathbf 0$ and projecting
onto a noncritical mode gives
$
\langle\mathbf v_n|\partial_i\mathbf K|\mathbf v_0\rangle
=-\lambda_n^{(N)}\langle\mathbf v_n|\partial_i\mathbf v_0\rangle,
$
the analogue of Eq.~\eqref{eq:eigvec_derivative}; its derivation used only
$\lambda_0^{(N)}=0$ and the Hermiticity of the static kernel, both of which hold
at $T_c$. Writing
\begin{equation}
    g_{0\mathbf{Q}}^{(N) n ij}
    =2\operatorname{Re}\left[
    \langle\partial_{Q_i}\mathbf v_0(\mathbf Q)|\mathbf v_n(\mathbf Q)\rangle
    \langle\mathbf v_n(\mathbf Q)|\partial_{Q_j}\mathbf v_0(\mathbf Q)\rangle
    \right]
    \label{eq:g0n_normal}
\end{equation}
for the pair quantum metric of Eq.~\eqref{eq:g0n} built from the eigenvectors
$\mathbf v_n(\mathbf Q)$ of $\mathbf K(\mathbf Q)$ rather than those of
$\mathbf M(\mathbf Q)$, the gradient coefficient becomes
\begin{equation}
    c_{ij}=\langle\mathbf v_0|\partial_i\partial_j
    \mathbf K|\mathbf v_0\rangle
    -\sum_{n>0}\lambda_n^{(N)} g_{0\mathbf{0}}^{(N) n ij}.
    \label{eq:cij}
\end{equation}
The GL gradient coefficient therefore carries the same stiffness-weighted
pair-geometric correction as the ordered-state stiffness: Eq.~\eqref{eq:cij} has
the structure of Eq.~\eqref{eq:Q_metric} with the ordered-state kernel replaced
by the normal-state one and without the overall factor
$\tfrac12$, so that
\begin{equation}
    c_{ij}
    =
    2\lim_{T\to T_c^-}\mathcal Q_{ij}(T).
    \label{eq:cij_Qij}
\end{equation}
As $T\to T_c^-$ the anomalous blocks of the ordered-state kernel vanish with the
order parameter, so that $\mathbf M(\mathbf Q)$ approaches the block-diagonal
form~\eqref{eq:M_normal} and the curvature of the Goldstone branch approaches
that of the critical normal-state eigenvalue; this is the precise sense in which
the two coefficients coincide
\footnote{
Unlike the ordered-state kernel, $\mathbf K(\mathbf Q)$ is not subject to
Eq.~\eqref{eq:ph_kernel}; however, in the presence of time-reversal symmetry
it obeys $\mathbf K(-\mathbf Q)=\mathbf K^*(\mathbf Q)$, which plays the same
role. When time-reversal symmetry is broken, the pair Berry curvature of
Sec.~\ref{sec:pairgeometry}, built from the eigenvectors
$\mathbf v_n(\mathbf Q)$ of $\mathbf K(\mathbf Q)$, therefore need not vanish
at $\mathbf Q=\mathbf 0$, and the isolated branches of $\mathbf K(\mathbf Q)$
may carry nonzero Chern numbers over the BZ of the pair momentum.
}.
In the remainder of this section $\mathcal Q_{ij}$
denotes this critical limit, $\mathcal Q_{ij}\equiv c_{ij}/2$.

The same limit can be taken term by term in Eq.~\eqref{eq:Ds_thermo}. Below
$T_c$ the critical mode acquires a uniform equilibrium amplitude $\Theta_0$,
with
$
\boldsymbol{\Delta}_0=\Theta_0\mathbf v_0+O(|\Theta_0|^3)
$
and hence
$
|\boldsymbol{\Delta}_0|^2=|\Theta_0|^2+O(|\Theta_0|^4)
$
since $\mathbf v_0$ is normalized.
For the block-diagonal kernel~\eqref{eq:M_normal} the Goldstone vector is
$
\mathbf w_0=(\ii\Theta_0\mathbf v_0,-\ii\Theta_0^*\mathbf v_0^*)^{\mathrm T}
/(\sqrt2|\Theta_0|),
$
and the two blocks contribute equally, so that the
normalization factor $\tfrac12$ cancels and
$
\langle\mathbf w_0|\partial_i\partial_j\mathbf M|\mathbf w_0\rangle
=\langle\mathbf v_0|\partial_i\partial_j\mathbf K|\mathbf v_0\rangle.
$
Equation~\eqref{eq:frozen_kernel} then gives
$
V[D_{\mathrm{frozen}}]_{ij}=4N_c|\Theta_0|^2\langle\mathbf v_0|
\partial_i\partial_j\mathbf K|\mathbf v_0\rangle
+O(|\Theta_0|^4).
$
In the same limit, each noncritical normal-state mode $\mathbf v_n$ generates
two massive modes in the doubled space, one from each block of
Eq.~\eqref{eq:M_normal}, both with stiffness
$\lambda_n\to\lambda_n^{(N)}$. Because the Goldstone vector carries equal
weight in the two blocks, each of these two modes contributes one half of
$g_{0\mathbf{0}}^{(N) n ij}$. Their sum therefore reproduces the corresponding
undoubled normal-state contribution in Eq.~\eqref{eq:relaxation_metric}, so that
$
\mathbf F_i^\dagger\boldsymbol{\mathcal H}(\mathbf 0)\mathbf F_j
=4|\Theta_0|^2\sum_{n>0}\lambda_n^{(N)} g_{0\mathbf{0}}^{(N) n ij}
+O(|\Theta_0|^4).
$
Combining the two contributions gives, to leading order near $T_c$,
\begin{equation}
    V[D_s]_{ij}=4N_c|\boldsymbol{\Delta}_0|^2c_{ij},
    \label{eq:Ds_cij}
\end{equation}
consistent with Eqs.~\eqref{eq:Ds_Qij} and \eqref{eq:cij_Qij}.

For the uniform critical pairing channel of the pyrochlore model of
Ref.~\cite{iskin24}, Eq.~\eqref{eq:Ds_cij} reproduces the near-$T_c$ relation
obtained there; here the critical mode need not be uniform across the
sublattices, and $c_{ij}$ then carries the pair-geometric term of
Eq.~\eqref{eq:cij}.

\subsection{Coherence length from the GL functional}
\label{sec:xi}

Projecting the static action onto the critical direction gives the
GL free-energy density per unit cell,
\begin{equation}
    f_{\mathrm{GL}}=a(T)|\Theta|^2+\frac{b_0}{2}|\Theta|^4
    +\sum_{ij}\mathcal Q_{ij} \nabla_i\Theta^*\nabla_j\Theta ,
    \label{eq:GL_functional}
\end{equation}
where $\Theta(\mathbf r)$ is the amplitude along $\mathbf v_0$, the real-space
counterpart of $\Theta(\mathbf Q)$ in the Fourier convention of
Sec.~\ref{sec:effaction}, with $\nabla_i\leftrightarrow \ii Q_i$, so that
the gradient term is the real-space form of the curvature
$\sum_{ij}\mathcal Q_{ij}Q_iQ_j$ of $K_{\mathrm{eff}}(\mathbf Q)$, and
$b_0>0$ is the coefficient of the quartic term generated by expanding 
the $\operatorname{Tr}\ln$ of Eq.~\eqref{eq:Seff} to fourth order in
$\boldsymbol{\eta}$ and projecting onto $\mathbf v_0$. The noncritical 
channels do not correct $b_0$ at this order. At $\mathbf Q=\mathbf 0$ 
and $T=T_c$ the off-diagonal blocks of Eq.~\eqref{eq:Gamma_eff} vanish, since
$\mathbf v_0$ diagonalizes the static kernel there, so that
$
a(T)=\langle\mathbf v_0|\mathbf K(\mathbf 0;T)|\mathbf v_0\rangle
$
and this coincides with $\lambda_0^{(N)}(T)$ to first order in $T-T_c$, the
drift of the critical eigenvector contributing only at second order. Writing
$\varepsilon=(T_c-T)/T_c$ and linearizing,
\begin{equation}
    a(T)\simeq-a_0\varepsilon,
    \qquad
    a_0=T_c \partial_T\lambda_0^{(N)}\big|_{T_c}>0 ,
    \label{eq:a0}
\end{equation}
minimization of Eq.~\eqref{eq:GL_functional} gives
$|\Theta_0|^2=a_0\varepsilon/b_0$, so that the superfluid weight vanishes linearly
at the transition,
$
V[D_s]_{ij}=4N_ca_0\varepsilon c_{ij}/b_0,
$
through the amplitude alone and not through the gradient coefficient.
Inserting a uniform twist,
$
\Theta=|\Theta_0|e^{\ii\phi}
$
with $\nabla_i\phi=2q_i$, into the gradient term of
Eq.~\eqref{eq:GL_functional} gives $V[D_s]_{ij}=4N_c|\Theta_0|^2c_{ij}$,
which is Eq.~\eqref{eq:Ds_cij} at the order to which it holds.

The coherence-length tensor follows from Eq.~\eqref{eq:GL_functional} without
reference to any mass scale, as the ratio of the gradient and quadratic
coefficients,
\begin{equation}
    [\xi^2_{\mathrm{GL}}]_{ij}=\frac{c_{ij}}{2a_0},
    \qquad
    [\xi^2(T)]_{ij} = [\xi^2_{\mathrm{GL}}]_{ij}/\varepsilon .
    \label{eq:xi}
\end{equation}
Because $c_{ij}$ carries the pair-geometric correction of Eq.~\eqref{eq:cij}, so
does the coherence length. The correction enters with a negative sign and $a_0$
is unaffected by it, so the relaxation of the noncritical channels shortens
every principal value of $\boldsymbol{\xi}_\mathrm{GL}$ relative to the value
obtained by holding the pairing configuration rigid across the sublattices.
This is the same softening that gives
$\mathbf{D}_s\preceq\mathbf{D}_\mathrm{frozen}$ below $T_c$: both quantities are
proportional to the relaxed curvature $\mathcal Q_{ij}$. When a single length is
wanted in place of the tensor, the natural choice is the geometric mean of its
principal-axis values,
$
\bar{\xi}_\mathrm{GL}
=\big[\det\boldsymbol{\xi}_\mathrm{GL}^{2}\big]^\frac{1}{2\bar d},
$
where $\bar d$ is the number of spatial dimensions in which the system is
extended; $\bar\xi_\mathrm{GL}$ is the length whose $\bar d$-dimensional
volume matches that of the coherence ellipsoid. Quantum-geometric contributions to the
coherence length of flat-band superconductors were discussed in the
uniform-pairing setting~\cite{iskin23, chen23, iskin24c}.

\subsection{Pair mass from the time-dependent GL theory}
\label{sec:TDGL}

The construction so far has used only the static kernel. To exhibit the
dynamics, we retain the frequency dependence of the normal-state propagator,
$\boldsymbol{\Gamma}^{-1}(\mathbf Q,\ii\nu_m)$, continue analytically to real
frequency, $\ii\nu_m\to\omega+\ii0^+$, and extend the kernel accordingly,
$
\mathbf K(\mathbf Q,\omega)\equiv
\boldsymbol{\Gamma}^{-1}(\mathbf Q,\omega+\ii0^+),
$
so that $\mathbf K(\mathbf Q,0)=\mathbf K(\mathbf Q)$. The dynamical
coefficient of the critical mode is
\begin{equation}
    d=-\partial_\omega
    \langle\mathbf v_0|\mathbf K(\mathbf 0,\omega)
    |\mathbf v_0\rangle\big|_{\omega=0, T_c} .
    \label{eq:d_coeff}
\end{equation}
The noncritical channels do not correct $d$. At $\mathbf Q=\mathbf 0$ the mixed
block $\mathbf K_{\perp0}(\mathbf 0,\omega)$ vanishes at $\omega=0$,
because $\mathbf v_0$ diagonalizes the static kernel; its leading frequency
dependence is therefore linear in $\omega$, and the Schur-complement correction
in Eq.~\eqref{eq:Gamma_eff} begins at order $\omega^2$. The pair-geometric
correction thus resides entirely in the gradient coefficient and not in the
dynamical one. Collecting the three coefficients, the effective kernel of the
critical mode has the time-dependent GL form
\begin{equation}
    K_\mathrm{eff}(\mathbf Q,\omega)
    =a(T)-d \omega+\sum_{ij}\mathcal Q_{ij}Q_iQ_j+\cdots ,
    \label{eq:TDGL}
\end{equation}
whose coefficients $a(T)$, $d$ and $\mathcal Q_{ij}$ are all defined directly
from the pair kernel.

Equations~\eqref{eq:Ds_Qij}, \eqref{eq:cij} and \eqref{eq:xi} require no
propagating pair mode, and do not depend on the interpretation that follows.
When the low-energy pair mode is well enough defined that $d$ is real and
positive, as it is when the mode lies below the two-particle continuum, the
canonically normalized field $\Theta_\mathrm{B}=\sqrt d \Theta$ obeys a
Gross-Pitaevskii equation whose inverse pair-mass tensor, pair repulsion,
chemical potential and condensate density are~\cite{sademelo93, iskin23}
\begin{equation}
\begin{aligned}
    [m_\mathrm{B}^{-1}]_{ij}&=\frac{c_{ij}}{d},
    &\qquad
    U_\mathrm{BB}&=\frac{b_0}{d^2},
    \\
    \mu_\mathrm{B}&=\frac{a_0\varepsilon}{d},
    &\qquad
    n^c_\mathrm{pair}&=\frac{N_c d |\Theta_0|^2}{V} .
    \end{aligned}
\label{eq:pairmass}
\end{equation}
The quartic coefficient $b_0$ of Eq.~\eqref{eq:GL_functional} thus appears as a
repulsive contact interaction between pairs, and the Gross-Pitaevskii equation
is completely specified by the coefficients of Eqs.~\eqref{eq:GL_functional} and
\eqref{eq:TDGL}. The identification is internally consistent: at the equilibrium
amplitude
$
|\Theta_{\mathrm{B},0}|^2=d|\Theta_0|^2=d a_0\varepsilon/b_0
$
one finds
$
U_\mathrm{BB}|\Theta_{\mathrm{B},0}|^2=a_0\varepsilon/d=\mu_\mathrm{B},
$
as a weakly interacting Bose condensate requires. For a single band this reduces to
the pair mass of Ref.~\cite{sademelo93}. In this regime, and only in this
regime, Eq.~\eqref{eq:Ds_cij} takes the Bose form
$
V[D_s]_{ij}=4Vn^c_\mathrm{pair}[m_\mathrm{B}^{-1}]_{ij}
$
near $T_c$, the factor of $4$ being the pair charge
squared~\cite{iskin19, iskin24}. In the dilute limit, where
$\boldsymbol{\Gamma}^{-1}(Q)$ reduces to the inverse vacuum two-body
$\mathcal T$ matrix, $[m_\mathrm{B}^{-1}]_{ij}=c_{ij}/d$ coincides with
the exact effective mass of the two-body bound state on a multiband lattice
obtained in Ref.~\cite{keskiner26}, including its pair-geometric contribution.

In the same reading, the coherence length of Eq.~\eqref{eq:xi} is the healing
length of this Bose gas~\cite{leggett}. Writing the standard Gross-Pitaevskii
relation
$
[\xi_\mathrm{B}^2(T)]_{ij}
=[m_\mathrm{B}^{-1}]_{ij}/2\mu_\mathrm{B}
$
in terms of its temperature-independent amplitude,
$
[\xi_\mathrm{B}^2(T)]_{ij} =[\xi_\mathrm{B}^2]_{ij}/\varepsilon,
$
gives
\begin{equation}
    [\xi_\mathrm{B}^2]_{ij}
    =\varepsilon \frac{[m_\mathrm{B}^{-1}]_{ij}}{2\mu_\mathrm{B}}
    =\frac{c_{ij}}{2a_0},
    \label{eq:xi_healing}
\end{equation}
which is the first of Eqs.~\eqref{eq:xi}. The dynamical coefficient cancels
between the inverse mass and the chemical potential,
which is why the coherence length follows from the static kernel
alone while the pair mass does not: only the ratio $c_{ij}/a_0$ is observable
in $\boldsymbol{\xi}_\mathrm{B}$, however heavy the pair.

Two remarks attach to Eq.~\eqref{eq:pairmass}. First, $n^c_\mathrm{pair}$ is
not a condensate density in the sense of off-diagonal long-range order but
the normalization that the Gross-Pitaevskii form imposes on the pairing
field; the canonical factor $\sqrt d$ is what makes the entries of
Eq.~\eqref{eq:pairmass} individually independent of the Hubbard-Stratonovich
normalization, unlike the product fixed by Eq.~\eqref{eq:Ds_Qij}. Second,
when low-energy quasiparticles are available at the energy of the critical
mode, $d$ acquires an imaginary part describing its
damping~\cite{sademelo93}, and the pair mass loses its meaning.

\section{Validity and scope}
\label{sec:validity}

Our derivation rests on the following assumptions: (i) the mean-field
saddle-point approximation and the Gaussian expansion of the pairing-field
action about it; (ii) a uniform saddle point, which in the twisted case means
uniform after the gauge transformation of Sec.~\ref{sec:thermo}; (iii) local
stability against static fluctuations at finite momentum,
$\mathbf M(\mathbf Q)\succeq0$ in a neighborhood of
$\mathbf Q=\mathbf 0$; (iv) a single soft static mode below $T_c$, namely the Goldstone
direction, with all other static eigenvalues positive; and (v) analyticity of
$\mathbf M(\mathbf Q)$ about $\mathbf Q=\mathbf 0$ with a
nondegenerate Goldstone branch, so that $\mathbf w_0(\mathbf Q)$ can be
continued smoothly. No time-reversal or inversion symmetry is assumed: below
$T_c$ the particle-hole structure~\eqref{eq:ph_kernel} supplies the evenness
of the Goldstone projection, and at $T_c$ only the stationarity of the
critical branch at $\mathbf Q=\mathbf 0$ is required (Sec.~\ref{sec:GL}).
Within these assumptions the channel structure is arbitrary.
Finite-range and interorbital pairing, spin-mixing spin-orbit coupling,
complex hoppings and coexisting mean fields only enlarge the set of pairing channels;
and if several continuous symmetries are broken simultaneously, the projection onto
$\mathbf w_0$ is replaced by a projection onto the soft subspace.

Note that $\mathcal Q_{ij}$ is defined from the static kernel alone
(Sec.~\ref{sec:TDGL}). In a system with gapless excitations the dynamical
kernel $\mathbf M(\mathbf Q,\ii\nu_m)$, after analytic continuation, contains
quasiparticle damping~\cite{sademelo93}, and the limits $\omega\to0$ and
$\mathbf Q\to\mathbf 0$ need not commute; the thermodynamic superfluid weight
is the static response, with the zero-frequency limit taken first. Gapless
fermions may also produce a nonanalytic momentum dependence of the static
kernel, which has to be checked case by case.

Assumption (i) is more basic. Beyond the Gaussian level the identification
$\mathbf M(\mathbf 0)=\boldsymbol{\mathcal H}(\mathbf 0)$ and the closed form in
terms of the mean-field amplitude are no longer available, although the
eigenvalue route of Eq.~\eqref{eq:Qij_soft_curvature} can still be applied to
the static pair kernel of whatever approximation is used.

\begin{figure*}[t]
    \centering
    \includegraphics[width=1\textwidth]{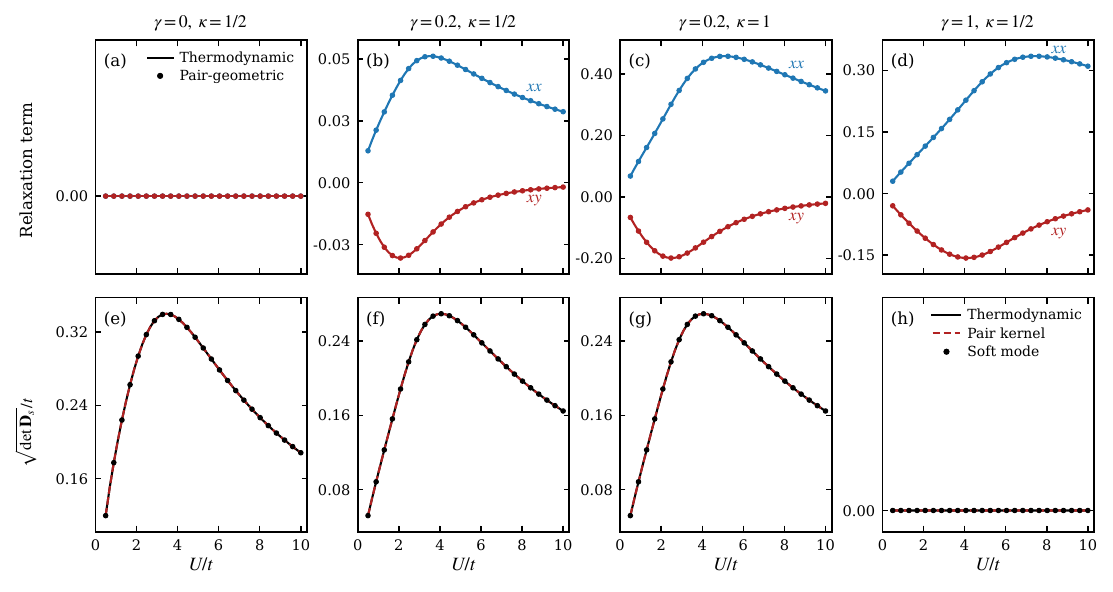}
    \caption{
    Numerical benchmark at $T=0$ as a function of the interaction strength $U/t$.
    The upper row, panels (a)--(d), compares the thermodynamic and pair-geometric
    representations of the relaxation term, shown in units of $t$. Solid lines show
    the components of $\mathbf F_i^\dagger\boldsymbol{\mathcal H}(\mathbf 0)\mathbf F_j$
    obtained from the self-consistent gap response, while filled circles show the
    stiffness-weighted pair quantum metric of Eq.~\eqref{eq:relaxation_metric}.
    Blue and red denote the $xx$ and $xy$ components, respectively.
    The lower row, panels (e)--(h), shows $\sqrt{\det\mathbf D_s}/t$ obtained from
    the thermodynamic formulation (black solid lines), the pair-fluctuation-kernel
    formulation (red dashed lines), and the curvature of $\lambda_0(\mathbf Q)$
    (black filled circles) according to Eq.~\eqref{eq:Qij_soft_curvature}.
    The columns correspond to (a,e) $(\gamma,\kappa)=(0,\tfrac12)$, (b,f) $(\gamma,\kappa)=(0.2,\tfrac12)$,
    (c,g) $(\gamma,\kappa)=(0.2,1)$, and (d,h) $(\gamma,\kappa)=(1,\tfrac12)$.
    }
    \label{fig:numerical_benchmark}
\end{figure*}
\section{Numerical benchmark}
\label{sec:numerics}

We now benchmark numerically the central identities of
Secs.~\ref{sec:thermo}--\ref{sec:supfwe} for the two-dimensional Lieb lattice
at $T=0$~\cite{huhtinen22}: the thermodynamic relaxation term of
Eq.~\eqref{eq:Ds_thermo} against its pair-metric form,
Eq.~\eqref{eq:relaxation_metric}, and the superfluid weight of
Eq.~\eqref{eq:Ds_thermo} against Eqs.~\eqref{eq:Ds_Qij} and
\eqref{eq:Qij_soft_curvature}. All
three are relations at fixed temperature, so a benchmark at $T=0$ tests them
completely; what the temperature controls is the value of
$\mathcal Q_{ij}$ and of the prefactor, not the identities themselves.

The Lieb lattice has three orbitals $S=A,B,C$ per unit cell, at
$\mathbf x_A=(0,\kappa)$, $\mathbf x_B=(0,0)$ and $\mathbf x_C=(\kappa,0)$ in units
of the lattice constant, which we set to unity, where $\kappa$
parametrizes the orbital embedding and $\kappa=\tfrac12$ corresponds to the
geometric positions of the orbitals.
The nearest-neighbor hopping amplitudes are
staggered as $t_\pm=t(1\pm\gamma)$, with $t_+$ within the unit cell and $t_-$
between neighboring cells. In the orbital basis $(A,B,C)$, the corresponding Bloch Hamiltonian
$
\mathbf h_{\mathbf k \uparrow} = \mathbf h_{\mathbf k \downarrow}
$
is
\begin{equation}
    \mathbf h_{\mathbf k}=
    \begin{bmatrix}
        0 & f_y(\mathbf k) & 0 \\
        f_y^*(\mathbf k) & 0 & f_x^*(\mathbf k) \\
        0 & f_x(\mathbf k) & 0
    \end{bmatrix},
\end{equation}
where
$
f_x(\mathbf k)=t_+e^{-\ii k_x\kappa}+t_-e^{\ii k_x(1-\kappa)}
$
and
$
f_y(\mathbf k)=t_+e^{-\ii k_y\kappa}+t_-e^{\ii k_y(1-\kappa)}.
$
Thus $\gamma$ controls the hopping staggering, while
$\kappa$ changes the embedding without changing the single-particle spectrum, and is
therefore the parameter against which the invariance of Eq.~\eqref{eq:Ds_Qij}
discussed in Appendix~\ref{app:embedding} can be tested directly.
For each $U/t$ the pairing amplitudes of Eq.~\eqref{eq:Delta_def} are determined
self-consistently from Eq.~\eqref{eq:gap}.
The mirror symmetry $x\leftrightarrow y$ through the $B$ site relates the
$A$ and $C$ orbitals, so that $\Delta_{0A}=\Delta_{0C}$ and the zero-twist
pairing vector is
$
\boldsymbol{\Delta}_0=(\Delta_{0A},\Delta_{0B},\Delta_{0A})^{\mathrm T}.
$
We work over $0.5\le U/t\le 10$ for the four parameter sets
$(\gamma,\kappa)=(0,\tfrac12)$, $(0.2,\tfrac12)$, $(0.2,1)$ and $(1,\tfrac12)$. The
first is a high-symmetry reference point; the two cases at $\gamma=0.2$ probe the
dependence of the separate contributions on the orbital embedding at fixed
hopping staggering; and $\gamma=1$ is the disconnected limit, which provides a
parameter-free check on the complete superfluid weight. These are the
parameter sets of the Lieb-lattice benchmark of Ref.~\cite{huhtinen22}.

The upper row of Fig.~\ref{fig:numerical_benchmark} compares the two
representations of the relaxation term. As discussed in Sec.~\ref{sec:thermo},
the thermodynamic contribution
$
\mathbf F_i^\dagger\boldsymbol{\mathcal H}(\mathbf 0)\mathbf F_j
$
is the order-parameter relaxation term identified in Ref.~\cite{huhtinen22},
obtained there by solving the differentiated gap equation, whereas
Eq.~\eqref{eq:relaxation_metric} expresses the same quantity as a sum over
massive channels weighted by their static stiffnesses, with no reference to the
gap response at all. The two evaluations agree to numerical precision across the
whole interaction range for all four parameter sets. At the high-symmetry point
$(\gamma,\kappa)=(0,\tfrac12)$ the relaxation term vanishes, while for finite staggering it
is nonzero and depends strongly on the embedding.
It is worth noting that the gaps are unequal, $\Delta_{0A}\neq\Delta_{0B}$,
in every panel, including the one in which the relaxation vanishes:
equal gap magnitudes at zero twist are not needed for the decoupling
condition of Sec.~\ref{sec:supfwe}, which at $(\gamma,\kappa)=(0,\tfrac12)$
follows instead from the inversion symmetry about the $B$ site; for
$\gamma=0$ but $\kappa\neq\tfrac12$ the relaxation term is
nonzero~\cite{huhtinen22}.

The lower row of Fig.~\ref{fig:numerical_benchmark} shows $\sqrt{\det\mathbf D_s}$,
the geometric mean of the principal values of the superfluid-weight tensor,
obtained in the three ways. The three agree to numerical precision throughout.
Two features deserve comment. First, although the relaxation contribution differs
substantially between the embeddings $\kappa=\tfrac12$ and $\kappa=1$ at fixed $\gamma=0.2$,
as panels (b) and (c) of Fig.~\ref{fig:numerical_benchmark} show,
the superfluid weight is identical in the two cases. This is the embedding
independence emphasized in Ref.~\cite{huhtinen22}, and here it appears
as a consequence of Eq.~\eqref{eq:Qij_soft_curvature}: the direct curvature and
the pair-geometric term are separately embedding dependent, while the eigenvalue
whose curvature they combine to give is not (Appendix~\ref{app:embedding}).
The compensation that must be arranged term by term in the thermodynamic
formulation is automatic in the collective-fluctuation one.
Second, the disconnected limit $\gamma=1$ provides a check that requires no
comparison between methods. There $t_-=0$, the unit cells decouple, and no
current can flow, so the superfluid weight must vanish identically at every
$U/t$. The frozen-pairing contribution does not
vanish: holding the pairing configuration rigid under the twist assigns a finite
superfluid weight to a system of disconnected clusters. It is cancelled exactly,
at every interaction strength, by the stiffness-weighted pair quantum metric of
Eq.~\eqref{eq:relaxation_metric}, and panel~(h) of Fig.~\ref{fig:numerical_benchmark}
shows the cancellation to numerical precision. This is the most direct
demonstration that the relaxation term is not a small correction to be estimated
but a necessary part of the superfluid weight~\cite{huhtinen22}.

\section{Conclusion}
\label{sec:conclusion}

We showed that the thermodynamic formulation of the superfluid weight, in
which the self-consistent relaxation of the pairing amplitudes enters
through the Hessian of the thermodynamic potential~\cite{huhtinen22}, has
an equivalent description in terms of collective pairing fluctuations: the
Hessian is the static pair-fluctuation kernel at zero momentum, the
relaxation correction is the Schur complement of the massive pairing
channels at finite pair momentum, and the superfluid weight is the
small-momentum curvature of the Goldstone branch of the kernel.

This reformulation exposed a quantum-geometric structure. As the pair
momentum is varied, the Goldstone eigenvector traces a path through the
space of pairing configurations, and the relaxation correction is its
quantum metric weighted by the static stiffnesses of the massive channels,
a collective counterpart of the Bloch quantum metric of the band structure.
The same geometry enters the GL gradient coefficient and the
coherence-length tensor near $T_c$, and, once the frequency dependence of
the kernel is retained, the pair mass~\cite{keskiner26}. The Lieb-lattice
benchmark confirmed these identities and the embedding independence of the
total superfluid weight, including the exact cancellation of the spurious
frozen-pairing stiffness in the disconnected limit~\cite{huhtinen22}.

Natural directions for future work include systems with several soft
collective modes, and the role of the pair Berry curvature and of the
topology of the normal-state pair-fluctuation branches when time-reversal
symmetry is broken.

\begin{acknowledgments}
We acknowledge support from the U.S. Air Force Office of Scientific
Research (AFOSR) under Grant No.~FA8655-24-1-7391.
\end{acknowledgments}

\appendix

\section{Alternative derivation of Eq.~\eqref{eq:M_and_H}}
\label{app:MH}

Here we derive Eq.~\eqref{eq:M_and_H} directly from the effective action of
Eq.~\eqref{eq:Seff}, by evaluating the thermodynamic Hessian
$\boldsymbol{\mathcal H}(\mathbf 0)$ of Eq.~\eqref{eq:H_blocks} and the
Gaussian fluctuation kernel $\mathbf M(Q)$ of Eq.~\eqref{eq:S2} as second
derivatives of $\mathcal S_{\mathrm{eff}}$ from the same starting point.
It is convenient to split the action as
$
\mathcal S_{\mathrm{eff}}=\mathcal S_\Phi+\mathcal S_f,
$
where
$
\mathcal S_\Phi=\frac{1}{U}\sum_{Q S}|\Phi_S(Q)|^2
$
and
$
\mathcal S_f=-\operatorname{Tr}\ln\left[-\mathbf G^{-1}\right],
$
with $\mathbf G^{-1}[\Phi]$ given by Eq.~\eqref{eq:Ginv}.

The BdG Hamiltonian is linear in the doubled pairing vector
$\boldsymbol{\Psi}=(\boldsymbol{\Delta},\boldsymbol{\Delta}^*)^{\mathrm T}$,
$
\mathbf h_{\mathrm{BdG}}(\mathbf k,\mathbf 0)
=\mathbf h_{\mathrm{BdG}}^{(N)}(\mathbf k)
+\sum_{\alpha=1}^{2N_S}\Psi_\alpha\mathbf V_\alpha,
$
with
$
\mathbf V_\alpha=\frac{\partial\mathbf h_{\mathrm{BdG}}}
{\partial\Psi_\alpha},
$
where
$\mathbf h_{\mathrm{BdG}}^{(N)}(\mathbf k)
$
is the normal-state BdG matrix of Eq.~\eqref{eq:Ginv}. Labeling the first
$N_S$ Nambu-orbital states as
particle-like and the last $N_S$ as hole-like, on-site pairing gives
$
\mathbf V_S=|S\rangle\langle N_S{+}S|
$
and
$
\mathbf V_{N_S+S}=\mathbf V_S^\dagger
$
for $S=1,\ldots,N_S$: the pairing
vertices connect the particle and hole components associated with the same
orbital $S$, and they carry no $\mathbf k$ dependence. The doubling of the
pairing vector implies
$
\Psi_\alpha^*=\Psi_{\bar\alpha}
$
with
$
\bar\alpha=\alpha+N_S\ (\mathrm{mod}\ 2N_S),
$
and therefore also
$
\mathbf V_\alpha^\dagger=\mathbf V_{\bar\alpha}
$
and
$
\partial_{\Psi_\alpha^*}=\partial_{\Psi_{\bar\alpha}}.
$

For a static and spatially uniform field,
Eq.~\eqref{eq:Ginv} is diagonal in $k$ with blocks
$
\ii\omega_n\mathbf 1-\mathbf h_{\mathrm{BdG}}(\mathbf k,\mathbf 0),
$
so that
$
\partial\mathbf G^{-1}/\partial\Psi_\alpha=-\mathbf V_\alpha.
$
Using
$
\delta\operatorname{Tr}\ln\mathbf A=\operatorname{Tr}(\mathbf A^{-1}
\delta\mathbf A)
$
with $\mathbf A=-\mathbf G^{-1}$, and then
$
\partial\mathbf G=-\mathbf G(\partial\mathbf G^{-1})\mathbf G,
$
gives
$
\frac{\partial \mathcal S_f}{\partial\Psi_\alpha}
    =\operatorname{Tr}\left[\mathbf G \mathbf V_\alpha\right],
$
and
$
\frac{\partial^2 \mathcal S_f}
    {\partial\Psi_\alpha^* \partial\Psi_{\alpha'}}
    =\operatorname{Tr}\left[
    \mathbf G \mathbf V_\alpha^\dagger \mathbf G
    \mathbf V_{\alpha'}\right],
$
where the second follows from the first by differentiating once more and using
$
\partial_{\Psi_\alpha^*}=\partial_{\Psi_{\bar\alpha}}
$
together with
$
\mathbf V_{\bar\alpha}=\mathbf V_\alpha^\dagger.
$
The pairing contribution to the same derivative is immediate: 
for a uniform static configuration with arbitrary $\boldsymbol{\Psi}$,
$
\Phi_S(Q)=\sqrt{\beta N_c} \Delta_S\delta_{Q0},
$
so that
$
\mathcal S_\Phi=\beta N_c\sum_S|\Delta_S|^2/U,
$
and since $\boldsymbol{\Delta}$ and
$\boldsymbol{\Delta}^*$ are independent variables,
$
(\beta N_c)^{-1}\partial_{\Psi_\alpha^*}\partial_{\Psi_{\alpha'}}
\mathcal S_\Phi=\delta_{\alpha\alpha'}/U.
$
Because
$
\mathcal S_{\mathrm{eff}}=\beta N_c \Omega_0(\mathbf 0,\boldsymbol{\Psi})
$
for such configurations, the Hessian of Eq.~\eqref{eq:H_blocks} is obtained
by dividing the sum of the two contributions by $\beta N_c$ and evaluating
the result at the saddle $\boldsymbol{\Psi}=\boldsymbol{\Psi}_0$,
\begin{equation}
    [\mathcal H(\mathbf 0)]_{\alpha\alpha'}
    =\frac{\delta_{\alpha\alpha'}}{U}+\frac{1}{\beta N_c}
    \sum_k \operatorname{tr}\left[\mathbf G_0(k) \mathbf V_\alpha^\dagger
    \mathbf G_0(k) \mathbf V_{\alpha'}\right],
    \label{eq:app_H_microscopic}
\end{equation}
where
$
\mathbf G_0(k)=[\ii\omega_n\mathbf 1
-\mathbf h_{\mathrm{BdG}}(\mathbf k,\mathbf 0)]^{-1}
$
is the Nambu Green's function at the self-consistent saddle.

We next obtain $\mathbf M(Q)$ from the same action.
Writing
$
\mathbf G^{-1}=\mathbf G_0^{-1}+\delta\mathbf G^{-1}
$
and expanding the logarithm, the quadratic fermionic contribution is
$
\mathcal S_f^{(2)}=\tfrac12\operatorname{Tr}[\mathbf G_0\delta\mathbf G^{-1}
\mathbf G_0\delta\mathbf G^{-1}].
$
The factor $\tfrac12$ here comes from the
expansion of $\ln(1+x)$ and not from any spinor redundancy: every Grassmann
field appears exactly once in the minimal $2N_S$-component spinor $\mathbf C_k$,
so that the fermionic determinant gives
$-\operatorname{Tr}\ln[-\mathbf G^{-1}]$ with unit coefficient. By
Eq.~\eqref{eq:Ginv}, a fluctuation of four-momentum $Q$ enters as
$
\delta\mathbf G^{-1}=-\frac{1}{\sqrt{\beta N_c}}\sum_{Q \alpha}
\delta\Phi_\alpha(Q)\mathbf V_\alpha,
$
with $\delta\boldsymbol{\Phi}(Q)$ the doubled fluctuation field of
Sec.~\ref{sec:below_Tc}, and connects two propagators whose
four-momenta differ by $Q$. The pairing contribution quadratic in the
fluctuations is
$
\mathcal S_\Phi^{(2)}=U^{-1}\sum_{Q S}|\eta_S(Q)|^2
=\tfrac12\sum_Q\delta\boldsymbol{\Phi}^\dagger(Q)(U^{-1}\mathbf 1)
\delta\boldsymbol{\Phi}(Q),
$
where the last equality uses the doubling
constraint of Eq.~\eqref{eq:doubling_constraint}; hence
$
\mathbf M^{\Phi}=U^{-1}\mathbf 1=\boldsymbol{\mathcal H}^{\Phi},
$
and the two constructions already agree in this sector.
Comparing the fermionic part with Eq.~\eqref{eq:S2} then gives
\begin{equation}
    [M(Q)]_{\alpha\alpha'}=\frac{\delta_{\alpha\alpha'}}{U}+
    \frac{1}{\beta N_c}\sum_k\operatorname{tr}\left[\mathbf G_0(k)
    \mathbf V_\alpha^\dagger \mathbf G_0(k+Q) \mathbf V_{\alpha'}\right].
    \label{eq:app_M_microscopic}
\end{equation}
Equations~\eqref{eq:app_H_microscopic} and \eqref{eq:app_M_microscopic} have the
same structure, and differ only through the argument of the second propagator.
At $Q=(\mathbf{0},0)$ the two Green's functions carry the same four-momentum,
Eq.~\eqref{eq:app_M_microscopic} reduces to Eq.~\eqref{eq:app_H_microscopic},
and Eq.~\eqref{eq:M_and_H} follows; Eq.~\eqref{eq:MQ_explicit} is the static
limit of Eq.~\eqref{eq:app_M_microscopic} after the Matsubara sum.

\section{Orbital-embedding dependence}
\label{app:embedding}

Here we show how a change of orbital embedding representing the same real-space
Hamiltonian is implemented in momentum space, how it transforms the Bloch
Hamiltonian and the static fluctuation kernel, and why the soft-mode curvature
$\mathcal Q_{ij}$ is invariant under it while the two contributions of
Eq.~\eqref{eq:Q_metric} separately are not. The embedding phases below depend on
the pair momentum alone, so the Matsubara frequency plays no role and we work
throughout with the static kernel $\mathbf M(\mathbf Q)$ of
Sec.~\ref{sec:below_Tc}.

We consider a change of orbital positions
$\mathbf x_S\rightarrow\mathbf x_S+\mathbf d_S$ at fixed real-space hopping
amplitudes, so that
$
\mathbf r_{lS}\rightarrow\mathbf r_{lS}+\mathbf d_S.
$
The real-space operators $c_{lS\sigma}$ are unchanged; what changes is the Fourier
convention of Sec.~\ref{sec:thermo}, which carries the orbital positions
explicitly. Equating the two expansions of the same operator,
$
\sum_{\mathbf k}e^{\ii\mathbf k\cdot\mathbf r_{lS}}c_{\mathbf kS\sigma}
=\sum_{\mathbf k}e^{\ii\mathbf k\cdot(\mathbf r_{lS}+\mathbf d_S)}
c'_{\mathbf kS\sigma},
$
gives
$
c'_{\mathbf kS\sigma}=e^{-\ii\mathbf k\cdot\mathbf d_S}c_{\mathbf kS\sigma}.
$
Introducing the diagonal unitary matrix
$
[W_{\mathbf d}(\mathbf k)]_{SS'}
=e^{-\ii\mathbf k\cdot\mathbf d_S}\delta_{SS'},
$
this reads
$
\mathbf c'_{\mathbf k\sigma}
=\mathbf W_{\mathbf d}(\mathbf k)\mathbf c_{\mathbf k\sigma},
$
and the Bloch Hamiltonian therefore transforms as
\begin{equation}
    \mathbf h'_{\mathbf k\sigma}
    =\mathbf W_{\mathbf d}(\mathbf k) \mathbf h_{\mathbf k\sigma}
    \mathbf W_{\mathbf d}^{\dagger}(\mathbf k),
    \label{eq:app_bloch_embedding}
\end{equation}
so that its spectrum is unchanged.
The on-site pair operator
$
b^\dagger_{S\mathbf Q}=\sum_{\mathbf k}
c^\dagger_{\mathbf k+\mathbf Q,S\uparrow}c^\dagger_{-\mathbf k,S\downarrow}
$
of Eq.~\eqref{eq:H_momentum} picks up the phases
$
e^{\ii(\mathbf k+\mathbf Q)\cdot\mathbf d_S}
$
and
$
e^{-\ii\mathbf k\cdot\mathbf d_S}
$
from its two factors. The $\mathbf k$
dependence cancels, as it does in Sec.~\ref{sec:thermo}, leaving
$
b_{S\mathbf Q}^{\prime\dagger}
=e^{\ii\mathbf Q\cdot\mathbf d_S}b_{S\mathbf Q}^{\dagger}
$
and
$
b'_{S\mathbf Q}=e^{-\ii\mathbf Q\cdot\mathbf d_S}b_{S\mathbf Q}.
$
Hence
$
b_{S\mathbf Q}^{\prime\dagger}b'_{S\mathbf Q}
=b_{S\mathbf Q}^{\dagger}b_{S\mathbf Q}
$
and the interaction term is invariant.
The embedding therefore changes the momentum-space representation of the
single-particle sector and the phase convention of the pair operators, while
leaving the full Hubbard Hamiltonian of Eq.~\eqref{eq:H_real} unchanged.

The Hubbard-Stratonovich field of Sec.~\ref{sec:effaction}, being conjugate to
$b_S(Q)$, transforms oppositely so as to preserve that coupling,
\begin{equation}
    \Phi'_S(Q)=e^{-\ii\mathbf Q\cdot\mathbf d_S}\Phi_S(Q),
    \label{eq:app_phi_embedding}
\end{equation}
so that in particular the uniform $\mathbf Q=\mathbf 0$ pairing field, and with
it the saddle $\boldsymbol{\Delta}_0$ of Eq.~\eqref{eq:Delta_def}, is unchanged.
For the doubled fluctuation field
$
\delta\boldsymbol{\Phi}(\mathbf Q)=(\boldsymbol{\eta}(\mathbf Q),
\boldsymbol{\eta}^*(-\mathbf Q))^{\mathrm T}
$
of Sec.~\ref{sec:below_Tc} both blocks transform with the same phase, since
Eq.~\eqref{eq:app_phi_embedding} gives
$
\eta_S^{\prime *}(-\mathbf Q)=e^{-\ii\mathbf Q\cdot\mathbf d_S}
\eta_S^*(-\mathbf Q).
$
Therefore
$
\delta\boldsymbol{\Phi}'(\mathbf Q)
=\boldsymbol{\mathcal W}_{\mathbf d}(\mathbf Q)\delta\boldsymbol{\Phi}(\mathbf Q)
$
with
$
\boldsymbol{\mathcal W}_{\mathbf d}(\mathbf Q)
=\operatorname{diag}[\mathbf W_{\mathbf d}(\mathbf Q),
\mathbf W_{\mathbf d}(\mathbf Q)].
$
Since
$
\boldsymbol{\mathcal W}_{\mathbf d}(\mathbf Q)
$
is unitary and the quadratic
action~\eqref{eq:S2} describes the same physical configuration in either
embedding, the corresponding kernels satisfy
\begin{equation}
    \mathbf M'(\mathbf Q)=\boldsymbol{\mathcal W}_{\mathbf d}(\mathbf Q)
    \mathbf M(\mathbf Q)
    \boldsymbol{\mathcal W}_{\mathbf d}^{\dagger}(\mathbf Q),
    \label{eq:embedding_similarity}
\end{equation}
a unitary similarity transformation at every $\mathbf Q$.
Let
$
\mathbf M(\mathbf Q)\mathbf w_n(\mathbf Q)
=\lambda_n(\mathbf Q)\mathbf w_n(\mathbf Q),
$
$n\ge0$, define the eigenvalue
branches of the static kernel, with $n=0$ the Goldstone branch of
Sec.~\ref{sec:pairgeometry}. Equation~\eqref{eq:embedding_similarity} implies
that
$
\boldsymbol{\mathcal W}_{\mathbf d}(\mathbf Q)\mathbf w_n(\mathbf Q)
$
is an eigenvector of $\mathbf M'(\mathbf Q)$ with the same eigenvalue, so that
$
\lambda'_n(\mathbf Q)=\lambda_n(\mathbf Q)
$
for every branch and every
$\mathbf Q$. In particular
$
\lambda'_0(\mathbf Q)=\lambda_0(\mathbf Q),
$
and Eq.~\eqref{eq:Qij_soft_curvature} then gives
\begin{equation}
    \mathcal Q'_{ij}
    =\frac{1}{2}\partial_i\partial_j\lambda'_0
    =\mathcal Q_{ij} .
    \label{eq:app_Q_invariant}
\end{equation}
The two terms of Eq.~\eqref{eq:Q_metric} are not individually embedding invariant.
To see this, write
$
\boldsymbol{\mathcal W}_{\mathbf d}(\mathbf Q)
=\exp\big(-\ii\sum_iQ_i\boldsymbol{\mathcal X}_i\big)
$
with
$
\boldsymbol{\mathcal X}_i=\operatorname{diag}
(d_{1i},\ldots,d_{N_S i}, d_{1i},\ldots,d_{N_S i}),
$
where $d_{Si}$ is the $i$th Cartesian component of $\mathbf d_S$.
Thus $\boldsymbol{\mathcal X}_i$ is the orbital-position
operator in the doubled pairing channels, and a change of embedding is the
$\mathbf Q$-dependent gauge transformation it generates, i.e.\ precisely the
kind of transformation under which a quantum metric is not invariant.
Since $\boldsymbol{\mathcal W}_{\mathbf d}(\mathbf 0)=\mathbf 1$, the
zero-momentum kernel and its eigenvectors are untouched:
$
\mathbf M'(\mathbf 0)=\mathbf M(\mathbf 0),
$
$
|\mathbf w'_n\rangle=|\mathbf w_n\rangle$ and $\lambda'_n=\lambda_n.
$
Differentiating Eq.~\eqref{eq:embedding_similarity} at $\mathbf Q=\mathbf 0$
gives
\begin{equation}
    \partial_i\mathbf M'=\partial_i\mathbf M
    -\ii\left[\boldsymbol{\mathcal X}_i,\mathbf M(\mathbf 0)\right],
    \label{eq:app_dM_embedding}
\end{equation}
while the second derivative acquires further terms involving
$\boldsymbol{\mathcal X}_i$ and $\boldsymbol{\mathcal X}_j$, so that the
direct-curvature term
$
\langle\mathbf w_0|\partial_i\partial_j\mathbf M|\mathbf w_0\rangle
$
changes. The pair-geometric term changes correspondingly: at finite momentum
$
|\mathbf w'_n(\mathbf Q)\rangle
=\boldsymbol{\mathcal W}_{\mathbf d}(\mathbf Q)|\mathbf w_n(\mathbf Q)\rangle
$
up to a $\mathbf Q$-dependent phase, which drops out of
Eq.~\eqref{eq:g0n}, and differentiating at $\mathbf Q=\mathbf 0$ gives
\begin{equation}
    |\partial_i\mathbf w'_n\rangle=|\partial_i\mathbf w_n\rangle
    -\ii\boldsymbol{\mathcal X}_i|\mathbf w_n\rangle ,
    \label{eq:app_dw_embedding}
\end{equation}
so that the mode-resolved metric $g_{0\mathbf 0}^{n ij}$ of Eq.~\eqref{eq:g0n}
is embedding dependent as well. The added term is not a pure phase, so neither
quantity is invariant on its own; only their combination in
Eq.~\eqref{eq:Q_metric}, which equals $\tfrac12\partial_i\partial_j\lambda_0$,
is, by Eq.~\eqref{eq:app_Q_invariant}. The Lieb-lattice benchmark of
Sec.~\ref{sec:numerics} shows both statements.

\section{Derivations of Eqs.~\eqref{eq:B_i} and \eqref{eq:frozen_kernel}}
\label{app:Bi}

Here we derive the relations between the phase-twist derivatives of the
thermodynamic formulation of Sec.~\ref{sec:thermo} and the momentum dependence
of the static fluctuation kernel of Sec.~\ref{sec:below_Tc}: the mixed
derivative $\mathbf B_i$ of Eq.~\eqref{eq:B_i} and the frozen-pairing
curvature of Eq.~\eqref{eq:frozen_kernel}.
We write $\mathbf r$ for the position of the pairing
component under consideration, with Cartesian components $r_i$, and we display
all sums over repeated Cartesian indices. With $\mathbf Q=2\mathbf q$ as in
Sec.~\ref{sec:supfwe}, every explicit $\mathbf q$ derivative at fixed
$\boldsymbol{\Psi}_0$ is a derivative along the winding
configuration~\eqref{eq:winding}.

At zero twist an infinitesimal global phase rotation of the uniform saddle
generates the Goldstone vector of Eq.~\eqref{eq:u0},
\begin{equation}
    \mathbf u_0=\ii\boldsymbol{\tau}_3\boldsymbol{\Psi}_0
    =\sqrt2|\boldsymbol{\Delta}_0|\;\mathbf w_0 ,
    \label{eq:app_goldstone}
\end{equation}
with $\mathbf M(\mathbf 0)\mathbf u_0=\mathbf 0$.
To linear order in the twist the phase-twisted
configuration is
$
\mathbf U_{\mathbf q}(\mathbf r)\boldsymbol{\Psi}_0=\boldsymbol{\Psi}_0
+2\sum_iq_ir_i\mathbf u_0+O(\mathbf q^2).
$
Let $\mathbf M[ \cdot ]$ denote the static kernel as a real-space operator,
i.e.\ the operator whose momentum representation is $\mathbf M(\mathbf Q)$,
with square brackets indicating the field it acts on; since it is the
Hessian of the static effective action at the saddle, the gradient generated by a small static variation
$
\delta\boldsymbol{\Phi}$ is $\mathbf M[\delta\boldsymbol{\Phi}]
$
to linear order. The coefficient of $q_i$ in the gradient on the twisted
configuration is therefore
$
\mathbf B_i=2\mathbf M[r_i\mathbf u_0].
$
To express this in momentum space we use
$
r_i\mathbf u_0
=-\ii\partial_{Q_i}[e^{\ii\mathbf Q\cdot\mathbf r}\mathbf u_0]_{\mathbf Q=\mathbf 0}
$
together with translational invariance,
$
\mathbf M[e^{\ii\mathbf Q\cdot\mathbf r}\mathbf u_0]
=e^{\ii\mathbf Q\cdot\mathbf r}\mathbf M(\mathbf Q)\mathbf u_0,
$
which gives
\begin{equation}
    \mathbf M[r_i\mathbf u_0]
    =-\ii \partial_{Q_i}\!\left[e^{\ii\mathbf Q\cdot\mathbf r}
    \mathbf M(\mathbf Q)\mathbf u_0\right]_{\mathbf Q=\mathbf 0}
    =-\ii(\partial_i\mathbf M)\mathbf u_0 ,
    \label{eq:app_Mr_u0}
\end{equation}
the term from differentiating the exponential being removed by
$\mathbf M(\mathbf 0)\mathbf u_0=\mathbf 0$. Hence
$
\mathbf B_i=-2\ii(\partial_i\mathbf M)\mathbf u_0,
$
which is Eq.~\eqref{eq:B_i}. The explicit factor of $\ii$ is what makes
$\mathbf B_i$ obey the same particle-hole condition as a gradient in
doubled pairing space, by the corollary
$
\boldsymbol{\tau}_1(\partial_i\mathbf M)^*\boldsymbol{\tau}_1
=-\partial_i\mathbf M
$
of Eq.~\eqref{eq:ph_kernel}.

We next relate the frozen-pairing
curvature to the direct Goldstone-sector curvature. Since the strictly linear
phase $\phi(\mathbf r)=2\mathbf q\cdot\mathbf r$ is not itself a periodic
fluctuation, we first extract the long-wavelength gradient coefficient using a
small periodic phase field, for which the Gaussian expansion is justified, and
only then evaluate it for the uniform twist. Consider
$
\boldsymbol{\Phi}(\mathbf r)=e^{\ii\phi(\mathbf r)\boldsymbol{\tau}_3}
\boldsymbol{\Psi}_0
$
with
$
\phi(\mathbf r_{lS})=\frac{1}{\sqrt{N_c}}\sum_{\mathbf Q}
e^{\ii\mathbf Q\cdot\mathbf r_{lS}}\phi(\mathbf Q).
$
For small $\phi$,
$
\delta\boldsymbol{\Phi}(\mathbf r)=\phi(\mathbf r)\mathbf u_0
-\phi^2(\mathbf r)\boldsymbol{\Psi}_0/2+\cdots$, where the $\phi^2
$
term multiplies the first variation of the action and therefore does not
contribute at the saddle. To quadratic order in $\phi$, Eq.~\eqref{eq:S2} gives
$
\mathcal S_{\mathrm{eff}}^{(2)}=\frac{\beta}{2}\sum_{\mathbf Q}|\phi(\mathbf Q)|^2
\langle\mathbf u_0|\mathbf M(\mathbf Q)|\mathbf u_0\rangle.
$
Expanding about $\mathbf Q=\mathbf 0$ gives
$
\langle\mathbf u_0|\mathbf M(\mathbf Q)|\mathbf u_0\rangle
=\tfrac12\sum_{ij}Q_iQ_j\langle\mathbf u_0|\partial_i\partial_j\mathbf M
|\mathbf u_0\rangle+O(\mathbf Q^4),
$
the zeroth-order term vanishing by the Goldstone condition and the linear
term by the evenness of
$
\langle\mathbf u_0|\mathbf M(\mathbf Q)|\mathbf u_0\rangle
$
in $\mathbf Q$ (Sec.~\ref{sec:lowQ}). Therefore
$
\mathcal S_{\mathrm{eff}}^{(2)}=\frac{\beta}{4}\sum_{ij}
\langle\mathbf u_0|\partial_i\partial_j\mathbf M|\mathbf u_0\rangle
\sum_{\mathbf Q}Q_iQ_j|\phi(\mathbf Q)|^2.
$
For the periodic test phase, Parseval's identity in the above convention gives
$
\sum_{\mathbf Q}Q_iQ_j|\phi(\mathbf Q)|^2
=\sum_l[\nabla_i\phi(\mathbf r_{lS})][\nabla_j\phi(\mathbf r_{lS})]
$
for any fixed orbital $S$. Using Eq.~\eqref{eq:app_goldstone} and
$\mathcal S_{\mathrm{eff}}=\beta N_c\Omega_0$ for a static configuration,
\begin{equation}
    \delta\Omega_0\big|_{\boldsymbol{\Psi}=\boldsymbol{\Psi}_0}
    =\frac{|\boldsymbol{\Delta}_0|^2}{2N_c}\sum_{ij}
    \langle\mathbf w_0|\partial_i\partial_j\mathbf M|\mathbf w_0\rangle
    \sum_l[\nabla_i\phi(\mathbf r_{lS})][\nabla_j\phi(\mathbf r_{lS})].
    \label{eq:app_dOmega}
\end{equation}
Having extracted the local gradient coefficient in Eq.~\eqref{eq:app_dOmega},
we evaluate it for the uniform twist, $\nabla_i\phi=2q_i$, for which the cell sum gives a factor $N_c$ and
$
\delta\Omega_0\big|_{\boldsymbol{\Psi}=\boldsymbol{\Psi}_0}
    =2|\boldsymbol{\Delta}_0|^2\sum_{ij}q_iq_j
    \langle\mathbf w_0|\partial_i\partial_j\mathbf M|\mathbf w_0\rangle.
$
Comparing with the quadratic expansion
$
\delta\Omega_0=\tfrac12\sum_{ij}
(\partial^2\Omega_0/\partial q_i\partial q_j)q_iq_j
$
yields
\begin{equation}
    \frac{\partial^2\Omega_0}{\partial q_i\partial q_j}
    \bigg|_{\mathbf q=\mathbf 0,
    \boldsymbol{\Psi}=\boldsymbol{\Psi}_0}
    = 4|\boldsymbol{\Delta}_0|^2
    \langle\mathbf w_0|\partial_i\partial_j\mathbf M|\mathbf w_0\rangle ,
    \label{eq:frozen_pairing_curvature}
\end{equation}
which is Eq.~\eqref{eq:frozen_kernel} after multiplication by $N_c$.
Equation~\eqref{eq:relaxation_metric} then follows by projecting
Eq.~\eqref{eq:hfc} onto the massive eigenvectors of
$\boldsymbol{\mathcal H}(\mathbf 0)=\mathbf M(\mathbf 0)$ and using
Eqs.~\eqref{eq:B_i}, \eqref{eq:app_goldstone} and
\eqref{eq:eigvec_derivative}, as done in Sec.~\ref{sec:supfwe}.

\bibliography{refs}

\end{document}